\documentclass[aps,prb,twocolumn,superscriptaddress,longbibliography]{revtex4-2}

\usepackage[dvipsnames]{xcolor}
\usepackage{bm}
\usepackage{graphicx}
\usepackage{upgreek}

\usepackage{natbib}
\usepackage{braket}
\usepackage{physics}
\usepackage{amssymb} 
\usepackage[%
breaklinks,%
pdftex,%
hyperfootnotes=true,%
pdfpagelabels,%
bookmarks,%
pageanchor,%
]{hyperref}

\hypersetup{%
	colorlinks=true, linktocpage=true, pdfstartpage=1, pdfstartview=FitH, pdfborder={0 0 0},%
	breaklinks=true, pdfpagemode=UseNone, pageanchor=true, pdfpagemode=UseOutlines,%
	plainpages=false, bookmarksnumbered, bookmarksopen=true, bookmarksopenlevel=1,%
	hypertexnames=true, pdfhighlight=/O,
	urlcolor=Red!50!Sepia, linkcolor=NavyBlue, citecolor=RoyalBlue, 
}
\newcommand{\ie}{i.e., } 
\def\>{\right\rangle}
\def\<{\left\langle}
\def\be{\begin{equation}}
\def\ee{\end{equation}}
\def\ba{\begin{array}{lll}}
\def\ea{\end{array}}

\def\beq{\begin{eqnarray}}
\def\eeq{\end{eqnarray}}
\def\d{{\rm d}}

\begin{document}
\title{Synchronization-induced violation of thermodynamic uncertainty relations}
\author{Luca Razzoli}
    \affiliation{Center for Nonlinear and Complex Systems, Dipartimento di Scienza e Alta Tecnologia, Universit\`a degli Studi dell'Insubria, Via Valleggio 11, 22100 Como, Italy} 
    \affiliation{Istituto Nazionale di Fisica Nucleare, Sezione di Milano, Via Celoria 16, 20133 Milano, Italy}
\author{Matteo Carrega}
\email{matteo.carrega@spin.cnr.it}
\affiliation{CNR-SPIN,  Via  Dodecaneso  33,  16146  Genova, Italy}
\author{Fabio Cavaliere}
    \affiliation{Dipartimento di Fisica, Universit\`a di Genova, Via Dodecaneso 33, 16146 Genova, Italy} 
\affiliation{CNR-SPIN,  Via  Dodecaneso  33,  16146  Genova, Italy}
		\author{Giuliano Benenti}
    \affiliation{Center for Nonlinear and Complex Systems, Dipartimento di Scienza e Alta Tecnologia, Universit\`a degli Studi dell'Insubria, Via Valleggio 11, 22100 Como, Italy} 
    \affiliation{Istituto Nazionale di Fisica Nucleare, Sezione di Milano, Via Celoria 16, 20133 Milano, Italy}
\author{Maura Sassetti}    
    \affiliation{Dipartimento di Fisica, Universit\`a di Genova, Via Dodecaneso 33, 16146 Genova, Italy} 
        \affiliation{CNR-SPIN,  Via  Dodecaneso  33,  16146  Genova, Italy}

\begin{abstract}
Fluctuations affect the functionality of nanodevices. Thermodynamic uncertainty relations (TURs), derived within the framework of stochastic thermodynamics, 
show that a minimal amount of dissipation is required to obtain a given relative energy current dispersion, that is, current precision has a thermodynamic cost. 
It is therefore of great interest to explore the possibility that TURs are violated, particularly for quantum systems, leading to accurate currents at lower cost.
Here, we show that two quantum harmonic oscillators are synchronized by coupling to a common thermal environment, at strong dissipation and low temperature.
In this regime, periodically modulated couplings to a second thermal reservoir, breaking time-reversal symmetry and taking advantage of non-Markovianity 
of this latter reservoir,  lead to strong violation of TURs for local work currents, while maintaining finite output power. 
Our results pave the way for the use of synchronization in the thermodynamics of precision. 
\end{abstract}
\maketitle

\section{Introduction}

In 1665 Huygens observed the synchronization of two pendulum clocks mounted on a common support~\cite{Huygens}.
Since then, synchronization has emerged as a universal concept in the theory of dynamical systems, with a broad range of applications in fields ranging from 
science and engineering to social life~\cite{Pikovsky}. More recently, the phenomenon has been investigated and characterized in quantum systems~\cite{Heinrich2011,zambrini_pra12, Giorgi2013, zambrini_scirep13,Lee2013,Walter2014,Bastidas2015,zambrini_pra16,du_scirep17,Witthaut2017,Amital2017,geng_jphyscom18,Roulet2018a,Roulet2018b,fazio_prr20}
with, however, only a few studies addressing thermodynamic signatures of synchronization~~\cite{Jaseem2020,Murtadho2023}.
Just as thermodynamics started in the 1800s spurred by the industrial revolution, in the same way the miniaturization of devices, and in particular the emergence of new quantum technologies, pushes the field of thermodynamics into new applied and fundamental challenges~\cite{espositormp, arrachea, sapienza, oro, lux24,arrachea1,rosa1}.
In the thermodynamics of small systems, fluctuations~\cite{bercioux15, dechiara, segal_fluct} play a prominent role, and thermodynamic uncertainty relations (TURs), derived within the framework of classical stochastic thermodynamics, 
establish a lower bound to the amount of dissipation needed to reduce relative energy current fluctuations to a given level~\cite{barato,shiraishi16, timpanaro19, horowitz17, pal20, proe, landi23, lou23, tesser23,sagawa23}. 
This seminal result motivated the quest for possible mechanisms to violate TURs, and consequently reduce the thermodynamic cost of precision.
Routes for TUR violations include breaking of time-reversal symmetry (TRS)~\cite{brandner18,cangemi_prr, taddei23} 
and quantum coherences~\cite{Ptaszynski2018,miller_prl_21, agarwalla,Liu2019,Guarnieri2019,cangemi_prb,Mitchison2021,Kalaee2021, saito23, paladino23, rosa23, rosa0, agar2, entropy22, muke22, acciai}. In addition, standard TURs have recently been generalized~\cite{cangemi_prr, brandner_unified, niggemann_2020}, including the case of time-dependent driving~\cite{koyuk_2019, koyuk_2019_1, koyuk_2020}. Although it might be intuitive that synchronization, by locking the relative motion of system constituents, can reduce fluctuations, its possible role in violating TURs has not yet been explored.

\begin{figure}[ht]
    	\centering
 \includegraphics[width=0.7\linewidth]{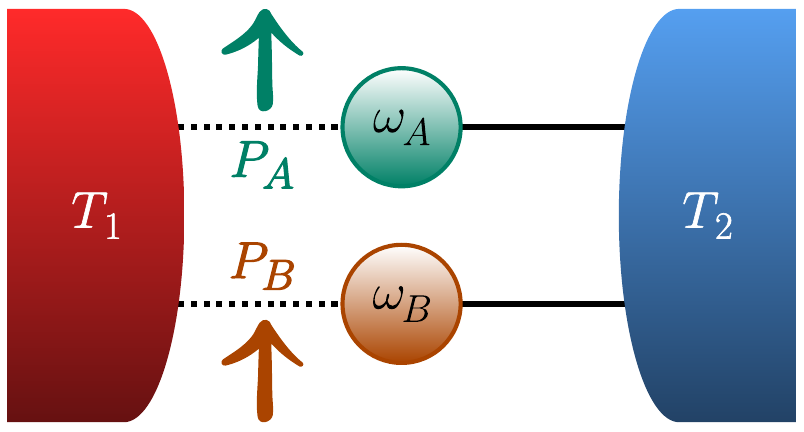}
\caption{\textbf{Sketch of the dynamical quantum thermal machine.}
Two quantum harmonic oscillators, with frequencies $\omega_A$ and $\omega_B$, are in contact with two common thermal reservoirs at temperatures $T_\nu$, with $\nu =1,2$. Power $P_{l}$ can be injected or extracted from the $l=A,B$ subsystem, while heat currents $J_\nu$ flow from or toward the reservoirs.
External monochromatic drives modulate the weak coupling with the $\nu=1$ reservoir, while the coupling with $\nu=2$ is static and much stronger, and it is responsible for synchronization.
\label{fig:1}}
\end{figure}

In this work, we consider two quantum harmonic oscillators (QHOs) coupled to common thermal baths (see figure~\ref{fig:1} for a schematic drawing
of our model). The couplings to one bath are static and induce not only dissipation but also the emergence of correlations between the two otherwise 
independent oscillators. At strong damping and low temperatures, the oscillators are synchronized, oscillating at a common frequency and in phase opposition. 
The oscillators are then in contact with another thermal bath, with periodically driven couplings breaking TRS. In this work, this simmetry breaking is the first necessary ingredient for TUR violation~\cite{barato}. 
We will also show that two other ingredients, namely synchronization and non-Markovianity, are required to achieve TUR violation together with finite output local power. Indeed, we will discuss that a strong violation for the injected or extracted power of each oscillator can occur, despite the fact that the TUR for total power is rigorously proven as never violated. 
In addition, the finite cutoff frequency for the dynamically coupled bath spectral density, which generally implies non-Markovian effects~\cite{frigerio, wc1, wc2}, allows local TURs to be violated in the fast driving diabatic regime.
Importantly, in this regime, the violation of TUR is accompanied by the possibility of extracting finite and  sizeable  power from one oscillator, exploiting synchronization achieved at strong damping. These results, already present in a isothermal regime, can benefit from the presence of a temperature gradient, especially in the non-linear regime where TUR violation is achieved in a wide parameter region. Finally, we show that local TURs can violate also the generalized bound for time-dependent drives~\cite{cangemi_prr, koyuk_2019, koyuk_2019_1, koyuk_2020}, even in the diabatic regime with finite output power. 

\section{General setting}
\subsection{Model} 

Two uncoupled (no direct coupling) quantum harmonic oscillators ($l=A,B$)  in contact with two common thermal reservoirs ($\nu=1,2$) are the subsystems constituting the working medium (WM) of the quantum thermal machine under study, as sketched in figure~\ref{fig:1}.
The total Hamiltonian is (we set $\hbar = k_B = 1$)
\begin{equation}
H^{(t)} = \sum_{l=A,B} H_{l} + \sum_{\nu=1,2} \big[ H_\nu + H_{{\rm int}, \nu}^{(t)}\big]~,
\label{eq:totH}
\end{equation}
with $H_l= \frac{p_l^2}{2 m} + \frac{1}{2}m \omega_l^2 x_l^2$ the Hamiltonian of the $l$-th QHO (same mass $m$ but different characteristic frequencies $\omega_A$ and $\omega_B$). The reservoirs $H_\nu$ are modelled with the Caldeira-Leggett approach of quantum dissipative systems~\cite{caldeira1983, weiss, weiss16, aurel18, dechiara_qho} as a collection of independent harmonic oscillators, while the system-reservoir interaction $H_{{\rm int},\nu}^{(t)}=\sum_{l=A,B} H_{{\rm int},\nu,l}^{(t)}$ is a bilinear coupling $\sim x_l \sum_{k=1}^{+\infty} g^{(l)}_\nu(t) c_{k,\nu} X_{k,\nu}$ in the subsystem $x_l$ and bath $X_{k,\nu}$ position operators, where $c_{k,\nu}$ describe the coupling strengths weighted by the modulating function $g^{(l)}_\nu (t)$ (see Appendix \ref{app_cl} for details). Notice that with the apex $^{(t)}$ we indicate the parametric time-dependence due to the presence of external drives.
We assume that the couplings with the $\nu=1$ reservoir are weak and oscillate in time~\cite{carrega_prxquantum, cavaliere_prr, cavaliere_iscience,carrega_arxiv}, with two independent monochromatic drives of the form
\begin{equation}
\label{eq:drives}
g_1^{(A)}(t)=f_A(\Omega )\cos(\Omega t)~,\quad
g_1^{(B)}(t)=f_B(\Omega) \cos(\Omega t+\phi)~,
\end{equation}
with $\Omega$ the external frequency, $\phi$ a relative phase, and $f_l(\Omega)$ relative weights that in general can depend on the external frequency $\Omega$. The couplings with the $\nu=2$ reservoir, instead, are static, $g_2^{(l)}=1$. Furthermore, the couplings with the $\nu=2$ bath are stronger than those with the $\nu=1$ bath. 

The properties of the $\nu$-th bath, including possible memory effects and non-Markovian behaviour~\cite{nm1,nm3,nm4,nm5}, are governed by the so-called spectral density~\cite{weiss} ${\cal J}_\nu(\omega)\equiv\frac{\pi}{2}\sum_{k=1}^{+\infty} \frac{c^{2}_{k,\nu}}{m_{k,\nu}\omega_{k,\nu}}\delta(\omega-\omega_{k,\nu})~$, where $m_{k,\nu}$ and $\omega_{k,\nu}$ are the mass and frequency of the $k$-th modes of the $\nu$-th bath.

As shown in Ref.~\cite{carrega_arxiv}, the out-of-equilibrium dynamics of the QHOs obey a set of coupled generalized quantum Langevin equations where the bath responses are encoded in the memory kernels
\begin{equation}
\label{eq:gamma}
\gamma_\nu(t)=\theta(t)\int_{-\infty}^{+\infty}\frac{\mathrm{d}\omega}{\pi m} \frac{{\cal J}_\nu(\omega)}{\omega}\cos(\omega t),
\end{equation}
with $\theta(t)$ the Heaviside step function, and a noise term, related to the fluctuating force $\xi_\nu(t)$ with null quantum average $\langle \xi_\nu(t)\rangle =0$ and correlation function~\cite{weiss, carrega_arxiv} 
\begin{align}
\label{xixi}
&\langle \xi_\nu(t) {\xi}_{\nu'}(t') \rangle = \delta_{\nu,\nu'}\int_0^\infty \frac{\d\omega}{\pi} {\cal J}_\nu (\omega)  \nonumber\\
& \times \left[ \coth\left(\frac{\omega}{2 T_\nu}\right) \cos[\omega (t-t')] - i \sin[\omega (t-t')]\right].
\end{align}

\subsection{Thermodynamic quantities}

In the following, we are interested in thermodynamic quantities in the long time limit, when a periodic steady state has been reached. 
To characterize the working regime and the performance of the quantum thermal machine, we focus on thermodynamic quantities averaged over the period ${\cal T}=2\pi/\Omega$ of the drives in the off-resonant case with $\omega_B < \omega_A$.
Due to the time-dependent drives in equation~\eqref{eq:drives}, power can be injected or extracted into/from the subsystems $l=A,B$. 
The average power associated to the $l$-th subsystem is defined as
\be
P_l \equiv \int_0^{{\cal T}}\frac{\d t}{{\cal T}} {\rm Tr}\left[\frac{\partial H_{{\rm int},1,l}^{(t)}}{\partial t}\rho(t)\right]~,
\label{eq:pldef}
\ee
where we have introduced both temporal and quantum averages, $\rho(t)$ is the total density matrix evolved at time $t$ (see Appendix \ref{app_cl}). 

The total power is given by $P=\sum_l P_l$. It is worth noting that power is associated to the temporal variation of the interaction term and, as such, is only due to the dynamical coupling of the WM to the bath $\nu = 1$. Furthermore, with our convention positive sign indicates power injection (or current flow toward the WM) and negative sign means power extraction (or current flow out of the WM).
The average heat current associated to the $\nu$-th reservoir is given by $J_\nu \equiv -\int_0^{{\cal T}}\frac{\d t}{{\cal T}}{\rm Tr}\big[{H}_\nu\dot{\rho}(t)\big]$, 
 and the balance relation $P+ J_1 + J_2=0$ holds true. We also recall that, in accordance with the second law of thermodynamics, the entropy production rate $\dot{S} \equiv -\sum_\nu J_\nu/T_\nu$ is always $\dot{S} \geq 0$~\cite{esposito2010, paternostro21}.

All quantities undergo fluctuations and the latter, once averaged over the period of the drives, can be written as~\cite{cangemi_prr, razzoli_epjst}

\be 
D_{O} \equiv \int_{0}^{{\cal T}}\ \frac{\d t}{{\cal T}}\int_0^{+\infty}\d\tau {\rm Tr}[\{O(t), O(t-\tau)\}\rho(t_0)]~,
\label{eq:corrOO}
\ee
for a generic operator $O(t)$, and where $\{\cdot,\cdot\}$ is the anticommutator. Notice that here the operators evolve in the Heisenberg picture with respect to the total Hamiltonian  and are at two different times.
Fluctuations, together with entropy production rate $\dot{S}$, are key figures of merit for thermal machines, which one often tries to minimize while having, e.g., finite power for a heat engine, in order to improve performance and stability of the thermal machine.
In particular, the impact of fluctuations on thermal machine performance can be assessed by standard TURs~\cite{cangemi_prr, barato, horowitz17, brandner18,Ptaszynski2018, agarwalla,Liu2019}.
The latter combine energy flows, their fluctuations, and the entropy production rate in a dimensionless quantity expressing the trade-off between the way the system fluctuates versus the quality (in terms of magnitude and degree of dissipation) of the energy flow. This trade-off parameter for a generic operator is given by
\begin{equation}
Q_{O}\equiv\dot{S}\frac{D_{O}}{O^2}~,
\label{eq:QP_def}
\end{equation}
and the standard TUR reads $Q_O\geq 2$.
Therefore, any mechanism leading to a violation of the TUR results in $Q_{O}<2$.
Among all the possible mechanisms for TUR violation we consider the breaking of TRS~\cite{brandner18, cangemi_prr}, 
which in our model is guaranteed by the presence of two independent drives with a finite phase shift $\phi \neq 0$. Notice that breaking of TRS is a necessary but not sufficient condition for TUR violation.
 Indeed, in the present setup it is possible to prove that the TUR for the total power $P$ is never violated, $Q_P \geq 2$ always, regardless of the value of $\phi$ and of the considered spectral densities (see Appendices \ref{app:der_pa_dpa} and \ref{app:QP_geq_2} for a proof of this result).
However, as we will show below, this is not the case 
for the properties of subsystems $l=A,B$, 
and TUR violation can be achieved by looking at $Q_{P_l}$ associated to the subsystem power.

\section{Synchronization}

\begin{figure}[ht]
    	\centering
 \includegraphics[width=\linewidth]{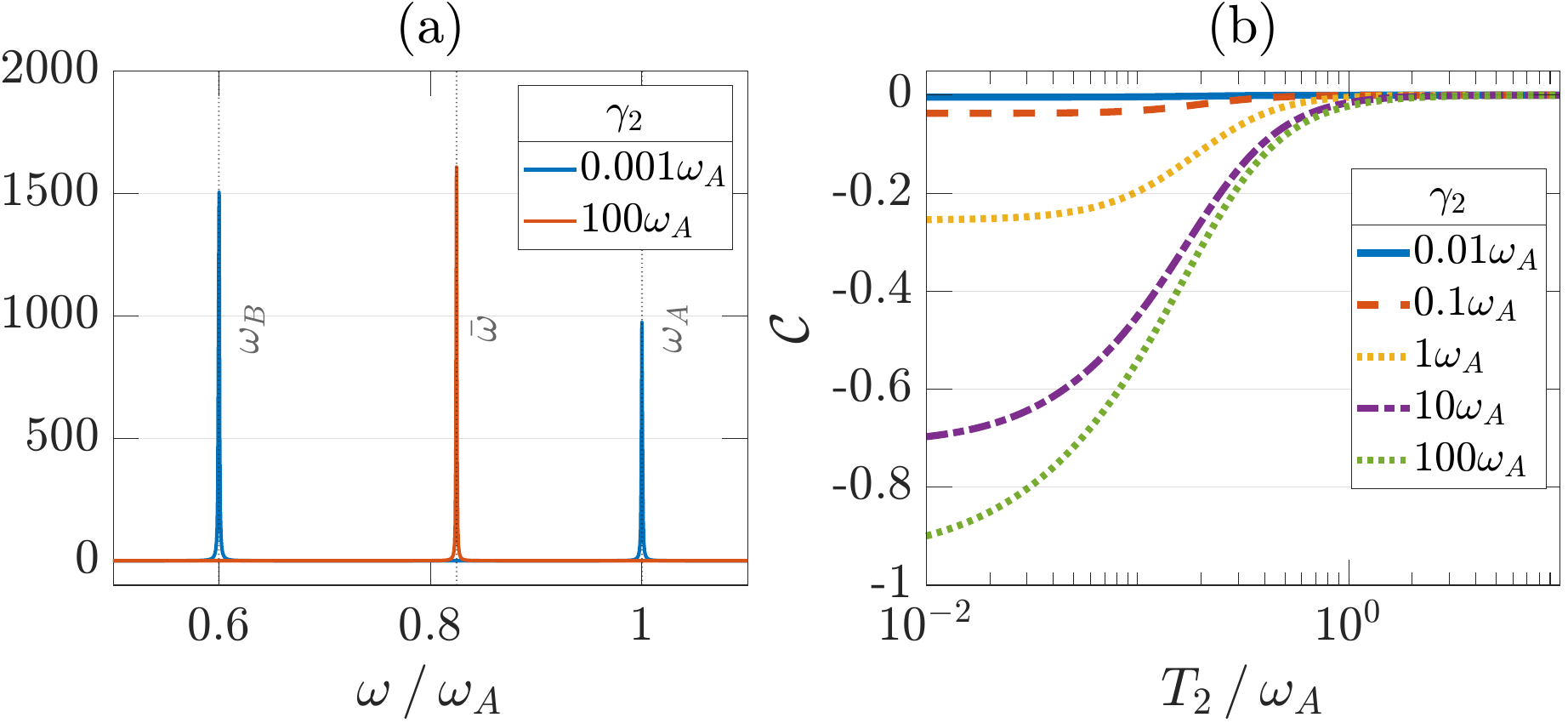}
\caption{\textbf{Strong damping regime and synchronization.} 
\textbf{(a)} Plot of the finite eigenvalue of ${\bm \chi}_2''(\omega)$ 
 as a function of frequency, showing that a dominant common mode appears at strong damping. \textbf{(b)} Synchronization, measured by the Pearson coefficient (\ref{eq:pearson}), as a function of temperature $T_2/\omega_A$. 
Different curves refer to various $\gamma_2$ damping strengths. Here and in the following figures,
$\omega_B=0.6\omega_A$.
\label{fig:1.5}}
\end{figure}

As stated above, we assume that the time-dependent couplings with the $\nu=1$ reservoir are much weaker than the static couplings with the $\nu=2$ one. Under this assumption, a systematic perturbative expansion governed by the ratio between the damping strengths $\gamma_1(\omega)/\gamma_2(\omega)$ (i.e., the Fourier transform of Eq.~\eqref{eq:gamma}) can be used, as formulated in Ref.~\cite{carrega_prxquantum}.
Therefore, the dynamics and all thermodynamic quantities will be evaluated at the lowest order in a perturbative expansion in the system-reservoir interaction $H_{{\rm int},1}^{(t)}$~\cite{cavaliere_prr, carrega_arxiv}. 
Conversely, the static couplings with the reservoir $\nu=2$ are treated to all order in the coupling strength. In the long time limit the static unperturbed problem is solved by ${\bf x}^\dag ={\bm \chi}_2 \cdot {\bm \xi }^\dag/m$, where ${\bf x} = (x^{(0)}_A(\omega) , x^{(0)}_B(\omega))$ is  the two-component vector of the positions of the oscillators, ${\bm \xi}=\xi_2(\omega) (1,1)$ is the noise vector and
${\bm \chi}_2(\omega)$ is the two-by-two response matrix, whose elements are the Fourier transform of the response function~\cite{carrega_arxiv}
\be
\label{eq:responsefunction}
\chi^{(l,l')}_2(t)\equiv i m \theta(t)\langle[x_l(t), x_{l'}(0)]\rangle~,
\ee
where $\langle \dots \rangle $ denotes the quantum average.
It is now worth to recall that in the resonant case ($\omega_A=\omega_B$) with static couplings to the $\nu=2$ reservoir, symmetry arguments lead to a dissipation-free subspace (associated to the relative coordinate normal mode $x_A-x_B$), preventing the system from reaching a stationary regime~\cite{carrega_arxiv, paz09, galve10, correa12, brenes}. For this reason, from now on we  consider $\omega_A \neq \omega_B$ only. Assuming a strictly Ohmic spectral density ${\cal J}_2(\omega)=m\gamma_2\omega$, the two-by-two response matrix is given by~\cite{carrega_arxiv}
\be 
\label{eq:chill}
\!\!\!\!\!\chi_2^{(l, l)}(\omega)\!=\!\frac{-[\omega^2-\omega_{\bar{l}}^2 + i \omega \gamma_2]}{{\cal D}(\omega)};\,
\chi_2^{(l, \bar{l})}(\omega)\!=\!\frac{i\omega\gamma_2}{{\cal D}(\omega)},
\ee
where we introduced the convention according to which if $l=A$ then $\bar{l} = B$ and \textit{vice versa}, and 
\be
\label{eq:denominator2}
{\cal D}(\omega)\!=\!(\omega^2\!- \omega_A^2)(\omega^2\!- \omega_B^2) + i\omega(2 \omega^2\! - \omega_A^2 - \omega_B^2)\gamma_2.
\ee
The response matrix is a key quantity since it determines the long-time dynamics and enters into the expressions of all the thermodynamic quantities of interest (see below). 
In particular, by inspecting the eigenvalue problem posed by ${\bm \chi}_2(\omega)$ (see above), at sufficiently strong damping $\gamma_2$ a frequency--  and phase--locked mode appears. Indeed, $\gamma_2$ not only determines dissipation, but it also mediates correlations between the two, otherwise independent, subsystems. There, the two subsystems $A$ and $B$ become synchronized, oscillating at a common frequency $\bar{\omega}=\sqrt{(\omega_A^2+\omega_B^2)/2}$ and in phase opposition. The appearance of a common frequency $\bar{\omega}$ can be inferred looking at the eigenvalues of the imaginary part of the response matrix of equation~\eqref{eq:chill} ${\bm \chi}_2(\omega)={\bm \chi}_2'(\omega)+i {\bm \chi}_2''(\omega)$, for different damping strengths $\gamma_2$. This is illustrated in figure~\ref{fig:1.5}(a) where the finite eigenvalue is reported for two different damping strengths. At weak damping $\gamma_2 \ll \omega_l$ two peaks around $\omega_A$ and $\omega_B$ are present, while at strong damping $\gamma_2 \gg \omega_l$ a unique common frequency at $\bar{\omega}$ is the dominant one. In this regime, the corresponding eigenvectors show that the two  QHOs are in phase opposition.

Mutual synchronization between the two QHOs arises when, regardless of their detuning, they start to oscillate coherently at a common frequency. 
In the present setup, therefore, synchronization can occur at strong damping $\gamma_2\gg \omega_l$.
To quantify synchronization, we consider the dynamics of local observables and the corresponding Pearson coefficient~\cite{qcorrsync_lecture}. It is worth to note that since we are interested in the steady state regime, the Pearson coefficient attains its asymptotic value and does not depend on time. 
Focusing on the position operator of the QHOs, this indicator reads
\be
\label{eq:pearson}
{\cal C}=\frac{\langle \delta x_A \delta x_B\rangle}{\sqrt{\langle \delta x_A\delta x_A\rangle \langle \delta x_B\delta x_B\rangle}}~,
\ee
where 
\begin{align}
\langle \delta x_l \delta x_{l'}\rangle & =  {\rm Tr}[x_l(t) x_{l'}(t)\tilde{\rho}] - {\rm Tr}[x_l(t)\tilde{\rho}]{\rm Tr}[x_{l'}(t)\tilde{\rho}]\nonumber \\
&= \int_{-\infty}^{+\infty}\frac{\d \omega}{4\pi m} \coth(\frac{\omega}{2T_2}) {\chi_2^{(l,l')}}''(\omega)~,
\end{align}
with $\tilde{\rho} = \exp\{-[\sum_{l=A,B} H_l + H_2 + H_{{\rm int},2,l}]/T_2\}$. 
The Pearson coefficient takes values between $-1$ and $+1$, respectively denoting perfect temporal anti-synchronization and synchronization of the local observables. The value $0$, instead, denotes the absence of synchronization~\cite{qcorrsync_lecture}. In figure~\ref{fig:1.5}(b) the behaviour of the Pearson coefficient for our setup with different values of the damping strength $\gamma_2$ is reported. This clearly shows 
 that at sufficiently strong damping the two subsystems become synchronized and in anti-phase, reaching ${\cal C}\to -1$ at low temperature. At high temperature this feature is smeared out, with 
Pearson coefficient ${\cal C}\to 0$. In the opposite, weak damping regime instead ${\cal C}$ is always small, indicating no synchronization.

\section{TUR violation for local power}

In the following, we will investigate the local powers and associated TURs~\cite{barato}. We will focus on the regime of strong damping and emphasize the role of synchronization 
to find parameter regions where useful, nonvanishing subsystem power 
($P_l <0$ and sizeable magnitude) can be obtained with high accuracy. 
We also recall that we consider a perturbative expansion in the ratios between the $\nu=1$ and $\nu=2$ system-bath coupling strengths. At the lowest perturbative order, the subsystem power contributions can be written as (see Appendix \ref{app:der_pa_dpa} for details)
\be
 P_l = P_l^{(0)} + \delta P_l,
\ee
with
\begin{align} 
& P_l^{(0)} = -\Omega \int_{- \infty}^{+ \infty} \frac{\d \omega}{4 \pi m}\mathcal{J}_1(\omega+\Omega) N(\omega,\Omega) \nonumber\\
& \times \left[ f_l^2(\Omega) {\chi_{2}^{(l,l)}}''(\omega) + \cos(\phi) f_l(\Omega)f_{{\bar l}}(\Omega) {\chi_{2}^{(l,{\bar l})}}''(\omega)\right], \label{eq:Pl0}\\
& \delta P_{A/B} \!=\!  \mp\Omega \sin(\phi) f_A(\Omega)f_B(\Omega) \nonumber\\
&\quad\quad \times \int_{- \infty}^{+ \infty}\!\!\! \frac{\d \omega}{4 \pi m}
\bigg[ {\cal J}_1(\omega+\Omega) {\chi_{2}^{(A,B)}}'\!(\omega) \coth\!\left(\! \frac{\omega\! +\! \Omega}{2T_1} \!\right) \nonumber\\
&\quad\quad - m(\omega\!+\!\Omega)\gamma_1''\!(\omega\!+\!\Omega) {\chi_{2}^{(A,B)}}''\!\!(\omega) \!\coth\!\left(\! \frac{\omega}{2T_2}\!\right)\!\!\bigg]\!, 
\label{eq:deltaPA}
\end{align}
where the last line involves the imaginary part of the damping kernel of equation~\eqref{eq:gamma} in Fourier space, $\gamma_1(\omega)=\gamma_1'(\omega)+i\gamma_1''(\omega)$, and we have introduced the function 
 $ N(\omega, \Omega) = \coth(\frac{\omega + \Omega}{2T_1}) -\coth(\frac{\omega}{2T_2})$. It is worth to stress that the total power is given by $P=P_A^{(0)} + P_B^{(0)}$, 
\begin{align}
\!&P = -\Omega \int_{- \infty}^{+ \infty} \!\frac{\d \omega}{4 \pi m}\mathcal{J}_1(\omega\!+\!\Omega) N(\omega,\Omega) \Big[f_A^2(\Omega) {\chi_2^{(A,A)}}''\!(\omega) \nonumber \\
\!&\!+\!f_B^2(\Omega){\chi_2^{(B,B)}}''\!(\omega)\!+\!2\cos(\phi)f_A(\Omega)f_B(\Omega){\chi_2^{(A,B)}}''\!(\omega)\Big]~.
\label{eq:PtotAB}
\end{align}
For the sake of completeness, we also report the expression for the $\nu=1$ heat current~\cite{carrega_arxiv}
\begin{align}\label{eq:j1exp}
\!& J_1 \!= \!\!\!\int_{- \infty}^{+ \infty} \!\!\!\frac{\d \omega}{4 \pi m}
(\omega\!+\!\Omega) \mathcal{J}_1(\omega\!+\!\Omega) N(\omega,\Omega) \Big[f_A^2(\Omega) {\chi_2^{(A,A)}}''\!(\omega) 
\nonumber \\
\!&\!+\!f_B^2(\Omega) {\chi_2^{(B,B)}}''\!(\omega)\!+\!2\cos(\phi)f_A(\Omega)f_B(\Omega) {\chi_2^{(A,B)}}''\!(\omega)\Big],
\end{align}
and $J_2= -P-J_1$. It is worth to underline that both $P$ and $J_\nu$ have an even dependence on $\phi$ and depend on the imaginary part of the response functions $\chi_2^{(l,l')}$ only. 
Importantly, the subsystem power contains a term $\delta P_l$ which is odd in the phase $\phi$. As we discuss in a moment, this will play a crucial role in determining the TUR violation for the subsystems $l=A,B$ since it represents an explicit TRS-breaking contribution. Indeed, setting $\phi=0$ TRS is not broken, $\delta P_l=0$, and no TUR violation occurs (see Appendix~\ref{app:QP_geq_2}). 
Finally, regarding the fluctuations associated to the subsystem power to the lowest perturbative order (see Appendix \ref{app:der_pa_dpa}) one gets
 \begin{align}
D_{P_l} = & \Omega^2 f_l^2(\Omega)\int_{-\infty}^{+\infty} \frac{\d \omega}{4 \pi m} \mathcal{J}_1(\omega+\Omega) N(\omega,\Omega)\nonumber\\
&\times \coth\left(\frac{\omega}{2T_2}-\frac{\omega+\Omega}{2T_1}\right) {\chi_2^{(l,l)}}''(\omega).
\label{eq:DPA}
\end{align} 
As already mentioned, we assume a strictly Ohmic spectral function for the $\nu=2$ reservoir, ${\cal J}_2(\omega)=m\gamma_2\omega$,
while the $\nu=1$ reservoir has an Ohmic spectral density in the Drude-Lorentz form
\be
\label{eq:drude}
{\cal J}_1 (\omega) = m\gamma_1 \frac{\omega}{1 +\frac{\omega^2}{\omega_{c}^2}}~,
\ee
with a cut-off frequency $\omega_c$.
Notice that in the strictly Ohmic regime, when $\omega_{c}$ is the highest energy scale, one recovers a memory-less (local in time) response.
Finite cut-off values $\omega_{c}$, instead, would in general imply non-Markovian effects~\cite{weiss, cattaneo1}.
It is worth to note that finite (and small) values of $\omega_{c}$ can be engineered in the context of quantum circuits~\cite{pekola11, cattaneo1, cattaneo2, esposito20} and have been already inspected in other related dissipative systems~\cite{frigerio, wc1, wc2, wc3}. With the spectral density of equation~\eqref{eq:drude} the imaginary part of the damping kernel $\gamma_1''(\omega)$ appearing in equation~\eqref{eq:deltaPA} becomes $\gamma_1''(\omega)={\cal J}_1(\omega)/(m\omega_c)$.

Hereafter we discuss in detail results for the $l=A$ channel and we exploit the $\phi$ phase degree of freedom, setting it to $\phi=\pi/2$ to maximize the TRS-breaking contribution. Furthermore we choose the relative weights as $f_A(\Omega)=1$ and $f_B(\Omega)= (\Omega/\omega_c)^r$ with the parameter $r\geq 0$. As we will see below this is a flexible choice that allows to obtain precise (small $Q_{P_A}$) but sizeable local power $P_A<0$, together with high efficiency $\eta$ at the same time. Note that analogous results can be obtained for $l=B$ letting $\phi = 3\pi/2$ and inverting the choice for the functions $f_l(\Omega)$.
  
Although the two subsystems $l=A,B$ have no direct coupling, correlations between the two are mediated by the interaction with the $\nu=2$ common reservoir, and we here consider the regime of full synchronization achieved at strong damping $\gamma_2\gg \omega_l$. Physically, this situation results in a very efficient exchange of power contributions between the two subsystems $P_A$ and $P_B$. This can be first illustrated in the case of isothermal reservoirs, $T_1=T_2=T$, where the two subsystems can act as a work-to-work converter~\cite{cangemi_prb}, 
e.g., $P_B\geq0$ is absorbed as input and $P_A \leq 0$ is extracted as output (with an associated efficiency $\eta = -P_A/P_B \leq 1$). The total power remains positive due to the second law of thermodynamics (indeed $\dot{S}=P/T\geq 0$).

By direct inspection of equation~\eqref{eq:DPA}, in the isothermal regime the fluctuations associated to $l=A$ reduce to $D_{P_A}=\Omega\coth(\Omega/(2T))P_A^{(0)}$, regardless of the shape of ${\cal J}_1(\omega)$, and 
\be
Q_{P_A} =  \frac{\Omega}{T} \coth\left (  \frac{\Omega}{2T}\right) \frac{P P_A^{(0)}}{P_A^2}
= Q_P \frac{1+ P_B^{(0)}/P_A^{(0)}}{(1+\delta P_A/P^{(0)}_A)^2},
\label{eq:QPA_isoT}
\ee
where $Q_P=\Omega\coth\left[\Omega/(2T)\right]/T$.
Since $Q_P \geq 2$ (see Appendix \ref{app:QP_geq_2}), violations of the TUR for $Q_{P_A}$, if any, must originate from the last fraction in the r.h.s. of the equation. In particular, one is interested in large values of the denominator, and hence one should look for $\delta P_A/P_A^{(0)}\gg 1$.

It is thus instructive to study the two opposite asymptotic behaviours of small and large external frequency. In the adiabatic regime ($\Omega \ll \omega_l, T$) one has
\beq
\label{eq:p_ad}
&&P_l^{(0)}\to \Omega^2  f_l^2(\Omega) \alpha_l^{({\rm ad})},\nonumber \\
&& \delta P_l \to \Omega f_B(\Omega) \delta \alpha^{({\rm ad})}_l~,
\eeq
where we have indicated with $\alpha_l^{({\rm ad})}$ and $\delta \alpha_l^{({\rm ad})}$ the expansion coefficients that do not depend on the external frequency $\Omega$ anymore (see Appendix \ref{app_asym} for their explicit expressions). 
This leads to the adiabatic expansion for the local TUR quantifier:
\be
\label{eq:qpa_ad}
 Q^{({\rm ad})}_{P_A} = 2 \Omega^2 \left(\alpha_A^{({\rm ad})}+f_B^2(\Omega) \alpha_B^{({\rm ad})}\right)\frac{\alpha_A^{({\rm ad})}}{f_B^2(\Omega){\delta \alpha_A^{({\rm ad})}}^2}~.
 \ee
In the opposite, diabatic, regime ($\Omega \gg \omega_l,\omega_c$) one obtains the following asymptotic expansions 
\beq
\label{eq:p_dia}
&& P_l^{(0)}\to \Omega {\cal J}_1 (\Omega) f^2_l(\Omega) \alpha_l^{({\rm dia})},\nonumber \\
&& \delta P_l \to \frac{\Omega^2}{\omega_c}f_B(\Omega) {\cal J}_1(\Omega) \delta \alpha_l^{({\rm dia})}~, 
\eeq
where again $\alpha_l^{({\rm dia})}$ and $\delta \alpha_l^{({\rm dia})}$ are frequency-independent expansion coefficients (see Appendix \ref{app_asym}).
These expressions depend on the shape of ${\cal J}_1(\Omega)$ and on the value of the cut-off $\omega_{c}$. The corresponding TUR quantifier becomes
\be
Q_{P_A}^{({\rm dia})} = \frac{\Omega}{T} \frac{1+f_B^2(\Omega)\alpha_B^{({\rm dia})}/\alpha_A^{({\rm dia})}}{\left[1+\tilde{{\cal C}} f_B(\Omega) \frac{\Omega}{\omega_c} \right]^2},
\label{eq:qpa_dia}
\ee 
where $\tilde{{\cal C}}= \delta \alpha_A^{({\rm dia})}/ \alpha_A^{({\rm dia})}$. 
Due to the above scaling behaviours, in both the asymptotic regimes one gets the possibility to achieve TUR violations as we will now discuss.

\subsection{Standard TUR violation}
\begin{figure}[ht]
    	\centering
 \includegraphics[width=\linewidth]{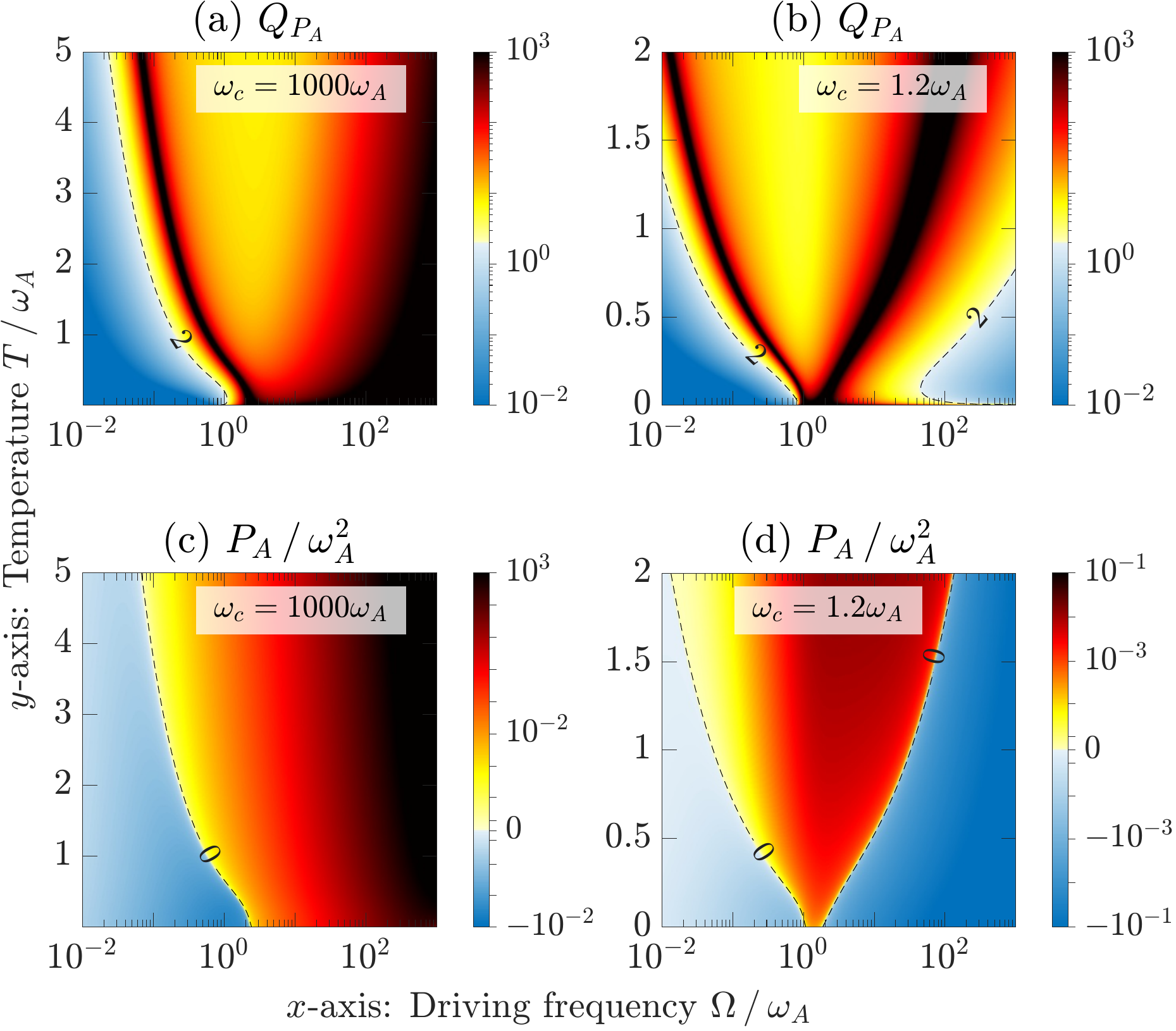}
\caption{\textbf{Quantifying fluctuations of $P_A$ via the trade-off quantity $Q_{P_A}$ in the isothermal regime.}
Density plots of $Q_{P_A}$ \textbf{(a)} and associated power $P_A$ \textbf{(c)} as a function of $\Omega/\omega_A$ and $T/\omega_A$ for $r=0$ in the large cut-off regime $\omega_{c}=1000\omega_A$. Panels \textbf{(b,d)} report the same quantities as in panels (a,c), respectively, but for a small cut-off $\omega_{c}=1.2\omega_A$. All results have been obtained in the strong damping regime $\gamma_2 =100\omega_A$; other parameters values are $\omega_B=0.6\omega_A$, $\gamma_1=0.01\omega_A$, and $\phi=\pi/2$. 
In panels (a,b) the dashed line denotes $Q_{P_A}=2$ separating regions where the TUR is violated (cool colors, $Q_{P_A}<2$) from regions where it is not (warm colors, $Q_{P_A}>2$). In panels (c,d) the dashed line refers to $P_A=0$ separating negative power contributions (power production, cool colors) from positive ones 
(power absorption, warm colors). 
\label{fig:2}}
\end{figure}

Let us now start by discussing possible violations from the standard TUR $Q_{P_A}\geq 2$.
To this end, we first consider the simple case of $r=0$ that corresponds to equal weigths $f_A(\Omega)=f_B(\Omega)=1$.
In the adiabatic regime, from Eq.~\eqref{eq:qpa_ad} one gets a $Q_{P_A}^{({\rm ad})}\propto \Omega^2$ regardless of the precise shape of the spectral density ${\cal J}_1(\omega)$.
This is indeed the case, as shown in figure~\ref{fig:2}(a-b)
There, density plots of $Q_{P_A}$ in the $\Omega-T$ plane are reported for two different values of $\omega_c$, where a region with $Q_{P_A} < 2$ is clearly visible in the left corner corresponding to $\Omega <\omega_A$. 
Looking at the corresponding power contribution $P_A$ (see panels c-d), one finds that $P_A <0$ in almost the same region, \ie the system is acting as a work-to-work converter.
 However, violation of the standard TUR in this parameter regime is associated with small power magnitude, with $P_A \to \Omega {\delta \alpha}^{({\rm ad})}_A$ in the adiabatic limit. In passing we mention that for the $l=B$ channel at $\phi=\pi/2$ analogous TUR violation are observed but always with $P_B >0$ values, that is $l=B$ cannot be used as a useful resource 
for power production (see Appendix \ref{app:pb}).

One may thus wonder if it is possible to achieve precise ($Q_{P_A}\ll 2$) but sizeable local power signals. To this end, one can look at the large frequency regime.
First of all, if one considers $\omega_{c}$ as the largest energy scale, thus with no memory effects, no TUR violations are expected (see the rightmost regions in the density plot of figure~\ref{fig:2}(a)). Intriguingly, the situation is different in the case of small cut-off $\omega_{c}$, \ie when non-Markovian effects become important. Indeed, in the case of small $\omega_c$ and considering finite values of $\tilde{{\cal C}}$ 
one thus gets $Q_{P_A}^{({\rm dia})}\propto 1/\Omega$, which again shows the possibility to get $Q_{P_A}< 2$. To corroborate this finding, in figure~\ref{fig:2}(b) the density plot of $Q_{P_A}$ in the $\Omega-T$ plane is reported for a representative small value $\omega_{c}=1.2\omega_A$. In this figure two regions where $Q_{P_A} < 2$ are present: the first in the adiabatic regime, as discussed above, and a second new region in the diabatic regime $\Omega \gg \omega_l,\omega_c$.
Importantly, this latter regime is also associated with finite power magnitude. This is indeed shown in figure~\ref{fig:2}(d), where a region with negative power contribution $P_A <0$  with sizeable magnitude is evident. Physically, in the diabatic regime $\Omega \gg \omega_l, \omega_{c}$, the external frequency is much higher than the cut-off 
frequency of the reservoir that becomes effectively freezed, and the large amount of injected power from the $l=B$ channel is almost entirely transferred to the $l=A$ one, resulting in a very efficient work-to-work conversion with efficiency $\eta\sim 1$ (see Appendix \ref{app:eta}).
 
\begin{figure}[!ht]
	\centering
	\includegraphics[width=\columnwidth]{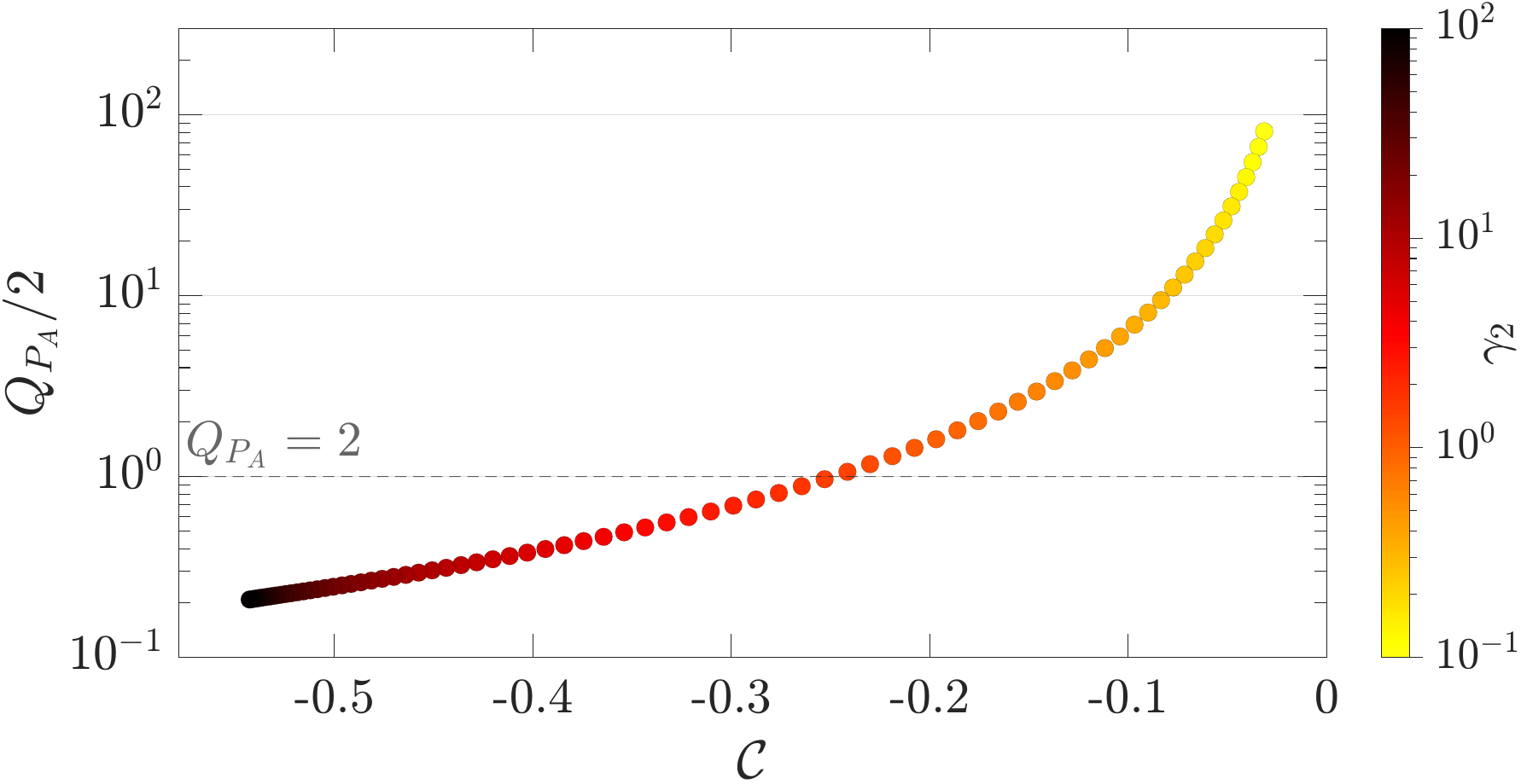}
	\caption{\textbf{Standard TUR violation and synchronization.} Parametric plot of the ratio $Q_{P_A}/2$ as a function of the Pearson coefficient at varying damping strengths $\gamma_2$. The stronger the damping, the more pronounced the (anti)synchronization ($\mathcal{C} \to -1$), the more the TUR is violated ($Q_{P_A}/2 < 1$). Results for $r=0$, $T = 0.1\omega_A$, $\omega_c = 1.2 \omega_A$, and $\Omega = 200 \omega_A$ (diabatic regime). Other parameters are $\omega_B = 0.6\omega_A$, $\gamma_1 = 0.01\omega_A$, and $\phi = \pi/2$.}
	\label{fig:tur_vs_pearson}
\end{figure}

We stress that in our model, in addition to
 TRS breaking, two are the key ingredients to achieve local TUR violation together with finite output power: memory effects and synchronization induced by strong damping $\gamma_2$. Indeed, related to the latter point, in Fig.~\ref{fig:tur_vs_pearson} one can notice a connection between the TUR quantifier and the Pearson coefficient ${\cal C}$. In particular, the stronger the synchronization, the lower the value of the TUR quantifier.
Interestingly, in the diabatic regime $\Omega \gg \omega_l, \omega_c$, a quantity reminiscent of the Pearson coefficient of equation~\eqref{eq:pearson} naturally appears. Indeed, looking at the denominator in equation~\eqref{eq:qpa_dia} and inspecting the definition of $\tilde{{\cal C}}$ one finds that
\be
\tilde{{\cal C}}=\frac{\langle \delta x_A\delta x_B\rangle}{\langle \delta x_A \delta x_A\rangle}= {\cal C}\sqrt{\frac{\langle \delta x_B\delta x_B\rangle}{\langle \delta x_A \delta x_A\rangle }}~, 
\ee
and it is $ |\tilde{{\cal C}}| < 1$. The behaviour of this Pearson-like coefficient is qualitatively the same as the one reported in figure~\ref{fig:1.5}(b) for the Pearson coefficient ${\cal C}$.
At low temperature $\tilde{{\cal C}}$ reaches values close to $-1$ in the strong damping regime when the two subsystems reach full synchronization being in anti-phase.
This allows to get large values of the denominator in equation~\eqref{eq:qpa_dia} that is a necessary condition to achieve TUR violation with sizeable power, clearly showing the importance of synchronization, established at strong damping.
In Appendix \ref{app_weakdamping} for the sake of completeness we have reported the 
behaviour of the TUR and the subsystem power in the case of weak damping $\gamma_2$, where synchronization is lacking, showing that there
the standard TUR for  ${P_A}$ is not violated in the diabatic regime.

\begin{figure}[ht]
    	\centering
 \includegraphics[width=\linewidth]{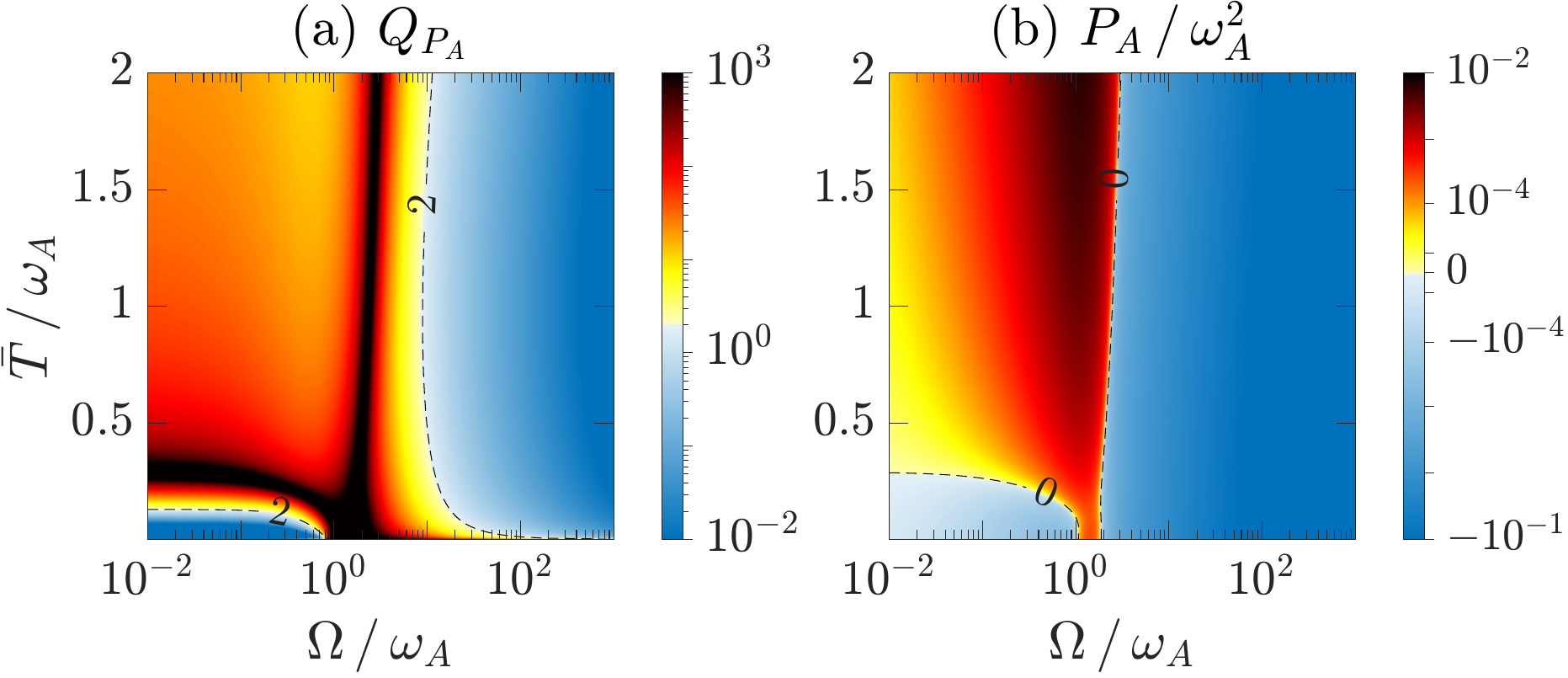}
\caption{\textbf{Quantifying fluctuations of $P_A$ via the trade-off quantity $Q_{P_A}$, in the presence of a temperature gradient in the strong damping regime $\gamma_2 = 100 \omega_A$, with cut-off $\omega_c=1.2\omega_A$ and for $r=0$.}
Density plots of \textbf{(a)} $Q_{P_A}$  and \textbf{(b)} $P_A$ as a function of the driving frequency $\Omega/\omega_A$ and average temperature $\bar{T}/\omega_A$ at relative temperature gradient $\Delta T / \bar{T} =1.8$. Other parameters as in figure \ref{fig:2}.
\label{fig:4}}
\end{figure}
So far we have discussed results in the isothermal regime with $T_1=T_2=T$. Before closing, we now demonstrate the robustness of our results in the presence of a relative temperature gradient $\Delta T / \bar{T}$, with $\bar{T}=(T_1+T_2)/2$ the average temperature and $\Delta T=T_1 - T_2$ the temperature difference. Not only the presence of a region of standard TUR violation with sizeable power in the diabatic regime still holds with a finite temperature gradient, but also wider regions are obtained in the non-linear regime of temperature gradient.
In figure \ref{fig:4}(a,b) we show results for $Q_{P_A}$ and $P_A$, respectively, as a function of $\Omega$ and $\bar{T}$, at given $\Delta T / \bar{T} = 1.8$. Notice that this last value implies a strong unbalance $T_2 \ll T_1$ (strongly non-linear regime).
At low driving frequencies, $\Omega < \omega_A$, the TUR for $P_A$ is violated for very low average temperature, $\bar{T}\lesssim 0.1\omega_A$. Instead, at higher driving frequencies the TUR is violated almost independently of the average temperatures considered, $0<\bar{T}<2\omega_A$. 
Comparing figure~\ref{fig:4}(b) and figure~\ref{fig:2}(b), we can see that the temperature gradient between the hot bath $\nu = 1$ and the cold one $\nu = 2$ 
can be exploited to greatly lower the driving frequency required to violate the TUR for $P_A$
(see also Appendix \ref{app_wth}). 
In closing this Section we mention that considering $r>0$ the standard TUR is still violated and no qualitative changes are found.

\subsection{Generalized TUR for periodic drives}
\begin{figure}[!ht]
	\centering
	\includegraphics[width=\columnwidth]{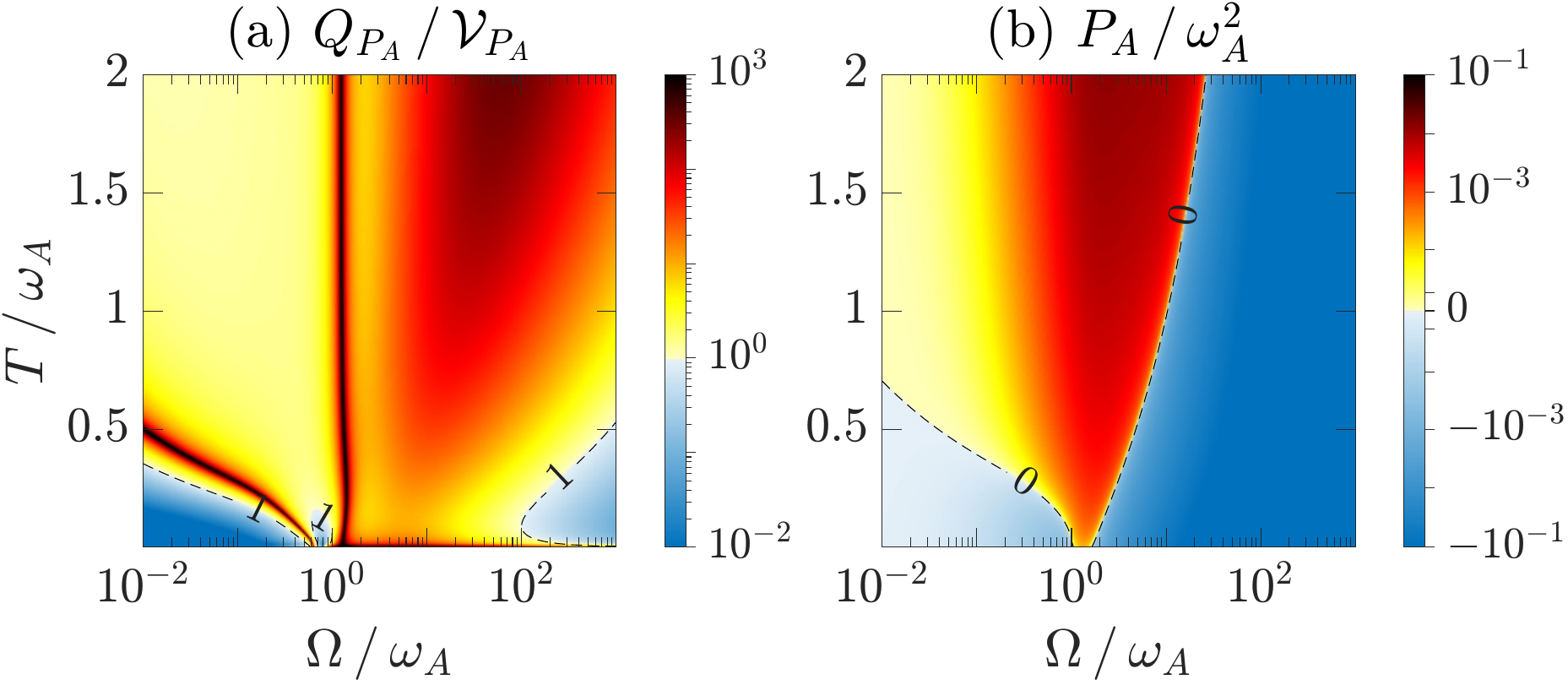}
	\caption{\textbf{Violation of the generalized TUR bound}. \textbf{(a)} Density plot of $Q_{P_A}/{\cal V}_{P_A}$ with $r = 1/2$ as a function of $\Omega/\omega_A$ and $T/\omega_A$ for small cut-off $\omega_{c}=1.2\omega_A$. Other parameters as in figure \ref{fig:2}. \textbf{(b)} Density plot of the local power $P_A$ for the same parameters as in Panel (a).}
	\label{fig:calv}
\end{figure}

So far we have discussed possible violations from the standard TUR bound $Q_{P_A}\geq 2$. However, the authors of Refs.~\cite{koyuk_2019, koyuk_2019_1, koyuk_2020}, considering a system under time-dependent drives, have derived a generalized bound $Q_O \geq {\cal V}_O$ with the quantity in the r.h.s. that in our case for the local power $P_l$ reads
\be
\label{eq:calv}
{\cal V}_{P_l}=2 \left(1-\Omega\frac{\partial_\Omega P_l}{P_{l}}\right)^2.
\ee
Focusing again on the $l=A$ channel, and looking for precise but sizeable output power, we now compare the TUR quantifier $Q_{P_A}$ to this generalized bound in the diabatic regime ($\Omega \gg \omega_l,\omega_c$).
First of all, inspecting Eq.~\eqref{eq:p_dia}, the asymptotic behaviour of Eq.~\eqref{eq:calv} in the case $r=0$ reads ${\cal V}_{P_A} \propto 1/\Omega^2$. Recalling that the TUR quantifier $Q_{P_A} \propto 1/\Omega$ in the diabatic regime, regardless of the value of the parameter $r$, the generalized TUR bound turns out to be satisfied with $Q_{P_A}\geq {\cal V}_{P_A}$.
However, one can consider a more general situation with $r>0$. Looking at Eq.~\eqref{eq:p_dia} this corresponds to optimize the magnitude of the local output power. In this case, the asymptotic behaviour of Eq.~\eqref{eq:calv} becomes ${\cal V}_{P_A} \to 2r^2$, independent of the external frequency $\Omega$.
This allows to look for the simultaneous desiderata of precise ($Q_{P_A}/{\cal V}_{P_A} \ll 1$), sizeable local power $P_A <0$, and high work-to-work conversion efficiency $\eta$ possibly close to unity in the isothermal regime. This last requirement imposes a constraint (see Appendix \ref{app:eta}) on the value of the parameter, leading to $0<r< 1$. Indeed, for $r>1$ the work to work efficiency in the diabatic regime tends to vanish. Therefore, by choosing $0<r< 1$ it is possible to obtain TUR violations, together with finite output power, even considering the generalized TUR bound posed by Eq.~\eqref{eq:calv}. In Fig.~\ref{fig:calv} we explicitly show a density plot of the TUR ratio $Q_{P_A}/{\cal V}_{P_A}$ for a representative value $r=1/2$, demonstrating the generalized TUR violation, again linked to the regime of synchronization achieved at strong damping, as discussed above.

\section{Discussion} 
We have shown that two otherwise independent quantum harmonic oscillators synchronize through coupling with a common thermal reservoir,
at strong dissipation and low temperature. When the oscillators are also dynamically coupled to a second thermal reservoir, 
the synchronization regime can be exploited to achieve strong TURs violation for local powers. Such violation exploits breaking of TRS 
by dynamical couplings. 
It is remarkable that, in the diabatic regime and when the dynamically coupled bath is non-Markovian, 
both the power and the amount of TUR violation increase with the driving frequency. 

Our results show the intimate connection between synchronization and thermodynamics of precision. 
From the standpoint of quantum technologies, synchronization mechanisms could be exploited to obtain finite and precise power
in quantum circuits, where non-Markovian environments can be engineered~\cite{pekola11, cattaneo1, cattaneo2, esposito20}.
From a more general point of view, a thermodynamic perspective could also be useful in the broad field of classical synchronization. These results open interesting perspectives for the exploitation of synchronization mechanisms, including bath-induced ones~\cite{new_arxiv}, beyond the paradigmatic model studied here. For instance it would be interesting to consider the impact of non-linearities and other systems including two-level systems that are commonly used in superconducting circuits.

%
%

\begin{acknowledgments}
M.C. and M.S. acknowledge support from the project PRIN 2022 - 2022PH852L (PE3) TopoFlags - "Non reciprocal supercurrent and topological transition in hybrid Nb-InSb nanoflags" funded within the programme "PNRR Missione 4 - Componente 2 - Investimento 1.1 Fondo per il Programma Nazionale di Ricerca e Progetti di Rilevante Interesse Nazionale (PRIN)", funded by the European Union - Next Generation EU. L.R. and G.B. acknowledge financial support from the Julian Schwinger Foundation (Grant JSF-21-04-0001) and from INFN through the project “QUANTUM”. L.R., F.C., and G.B. acknowledge support from the project PRIN 2022 - 2022XK5CPX (PE3) SoS-QuBa - "Solid State Quantum Batteries: Characterization and Optimization" funded within the programme "PNRR Missione 4 - Componente 2 - Investimento 1.1 Fondo per il Programma Nazionale di Ricerca e Progetti di Rilevante Interesse Nazionale (PRIN)", funded by the European Union - Next Generation EU.
\end{acknowledgments}

\appendix
\section{System-reservoir interactions}
\label{app_cl}
Here we provide some details of the microscopic model of the reservoirs and their couplings with the working medium of the quantum thermal machine.
Following the Caldeira-Leggett approach~\cite{weiss}, the $\nu$-th bath Hamiltonian reads
\be
H_{{\nu}}=  \sum_{k=1}^{+\infty} \left(\frac{P^2_{k,\nu}}{2 m_{k,\nu}} + \frac{1}{2}m_{k,\nu} \omega^2_{k,\nu }X^2_{k,\nu}\right).
\label{eq:Ham_res}
\ee
The bilinear form describing the WM-bath interactions is given by
\begin{align}
& H_{{\rm int},\nu}^{(t)} = \sum_{l=A,B} \sum_{k=1}^{+\infty} \bigg[
- g_\nu^{(l)}(t) c_{k,\nu} x_l X_{k,\nu}
\nonumber \\
&+\frac{(g_\nu^{(l)}(t) c_{k,\nu})^2}{2 m_{k,\nu}\omega_{k,\nu}^2} x_{l}^2 
+ \frac{g_\nu^{(l)}(t) g_\nu^{(\bar{l})}(t)  c_{k,\nu}^2 }{2 m_{k,\nu}\omega_{k,\nu}^2} x_{l}x_{\bar{l}} \bigg]~,
\label{eq:H_int_nu}
\end{align}
with  $c_{k,\nu}$ describing the coupling strengths between the QHOs and the $k$-th mode of the $\nu$-th reservoir, modulated by the drives $g_\nu^{(l)}(t)$. In the above equation we used  the convention according to which if $l=A$ then $\bar{l} = B$, and \textit{vice versa}. 
Note that the superscript $^{(t)}$ reminds the time-dependent modulation $g_1^{(l)}(t)$.
It is important to note that the interaction in equation~\eqref{eq:H_int_nu} includes counter-term contributions having a twofold purpose: (i) to avoid renormalization of the characteristic frequencies $\omega_{A,B}$ of the QHOs and (ii) to cancel the direct coupling between the latter that would naturally arise~\cite{carrega_arxiv}.

At the initial time $t_0{\to -\infty}$ we assume that the reservoirs are in their thermal equilibrium at temperatures $T_\nu$ and that the total density matrix is in a factorized form $\rho(t_0) = \rho_{A}(t_0)\otimes\rho_{B}(t_0)\otimes \rho_{1}(t_0)\otimes \rho_{2}(t_0)$, with $\rho_{l}(t_0)$ the initial density matrix of the $l$-th QHO, and $\rho_\nu (t_0) = \exp(-H_\nu / T_\nu)/\Tr[\exp(-H_\nu / T_\nu)]$ the thermal density matrix of the $\nu$-th reservoir.

\section{Derivation of power $P_A$ and associated fluctuations $D_{P_A}$}
\label{app:der_pa_dpa}
Here we provide details on the derivation of the subsystem average power $P_l$ and its fluctuations. We start from the definition in equation~\eqref{eq:pldef} and we first focus on the term present in the r.h.s.
\be
\langle P_l(t)\rangle= {\rm Tr}\bigg[\frac{\partial H_{{\rm int},1,l}^{(t)}}{\partial t}\rho(t)\bigg]~,
\ee
where we have indicated with $\langle \cdot \rangle$ 
the quantum average.
Denoting with $x_l^{(0)}(t)$ the unperturbed 
position operator, at the lowest perturbative order we obtain
\begin{align} 
& \langle P_l(t)\rangle \! = \! \dot{g}_1^{(l)}(t)\Bigg\{ \! i \! \sum_{l'}\!\! \int_{t_0}^{t} \!\!\!\! \d s \, {\cal L}_1 (s \!-\! t)g_1^{(l')}\!(s) \big\langle\! \big[ x_{l'}^{(0)}\!(s), x^{(0)}_{l}\!(t)\big] \!\big\rangle \nonumber \\
& +\sum_{l'}\int_{t_0}^{t} \d s \,\dot{{\Gamma}}_1(t \!-\! s) g_{1}^{(l')}\! (s) \big\langle x^{(0)}_l\! (t)x^{(0)}_{l'}\! (s) \big\rangle +g_1^{(l)}\! (t) {\Gamma}_1(0) \nonumber \\
&\times \!  \big\langle x_l^{(0)}\! (t)x_l^{(0)}\! (t)\big\rangle + g_1^{({\bar l})}\! (t){\Gamma}_1(0) \big\langle x_l^{(0)}(t)x_{{\bar l}}^{(0)}(t)\big\rangle \! \Bigg\}
\end{align}
where ${\Gamma}_1(t)$ is defined through equation~\eqref{eq:gamma} via $\gamma_1(t)=\frac{1}{m}\theta(t){\Gamma}_1(t)$. 
Here, we have introduced the symmetric and anti-symmetric contributions of the correlation function of equation~\eqref{xixi} $\langle\xi_\nu(t)\xi_{\nu'}(s)\rangle \equiv {\cal L}_\nu(t-s)\delta_{\nu,\nu'}$, with ${\cal L}_\nu(t-s) ={\cal L}_{\nu,{\rm s}} (t-s) + {\cal L}_{\nu,{\rm a}}(t-s)$ where
\begin{align}
{\cal L}_{\nu,{\rm s}}(t) &= \int_0^{+\infty} \frac{\mathrm{d}\omega}{\pi}
{\cal J}_\nu(\omega)\coth\left(\frac{\omega}{2T_\nu}\right)\cos(\omega t),\label{eq:ls}\\
{\cal L}_{\nu,{\rm a}}(t) &= -i\int_0^{+\infty} \frac{\mathrm{d}\omega}{\pi} 
{\cal J}_\nu(\omega)\sin(\omega t).\label{eq:la}
\end{align}
Changing variable $\tau=t-s$ and introducing $C^{(l,l')}(t-s)=C^{(l,l')}(t,s)\equiv \big\langle x_l^{(0)}(t)x_{l'}^{(0)}(s)\big\rangle$ we rewrite
\begin{align}
\!&\!\langle P_l(t)\rangle \! = \!  i  \sum_{l'}\! \int_{0}^{+\infty}\!\! \d \tau \, {\cal L}_1(-\tau)G^{(l,l')}(t,\tau)\Big[C^{(l',l)}(-\tau)  \nonumber \\
\!&\!-\! C^{(l,l')}(\tau)\Big] +\! \sum_{l'}\int_{0}^{+\infty} \!\! \d \tau \,\dot{\Gamma}_1(\tau) G^{(l,l')}(t,\tau) C^{(l,l')}(\tau)  \nonumber \\
\!&\!+G^{(l,l)}(t,0) \Gamma_1(0) C^{(l,l)}(0) + G^{(l,{\bar l})}(t,0)\Gamma_1(0) C^{(l,{\bar l})}(0), 
\end{align}
where ${\bar l}=B$ if $l=A$ (and viceversa), and we have introduced
\be
\label{g_grande}
G^{(l,l')}(t,\tau)= \dot{g}_1^{(l)}(t) g_1^{(l')}(t-\tau).
\ee
Recalling the identity
\be
\dot{\Gamma}_1(t)=-2i {\cal L}_{1,{\rm a}}(t)~,
\ee
introducing the average over the period and defining
\be 
G^{(l,l')}(\tau)=\int_{0}^{{\cal T}}\frac{\d t}{{\cal T}}G^{(l,l')}(t,\tau) 
\ee
and the combination of correlators
\be
\label{cpm}
C_\pm^{(l,l')}(\tau)= C^{(l,l')}(\tau) \pm C^{(l',l)}(-\tau)
\ee
we get
\begin{align}
\label{pa_1}
&P_l \!=\! -i \! \int_{0}^{+\infty}\!\!\!\!  \d \tau \, {\cal L}_{1,{\rm s}}(\tau)\Big[G^{(l,l)}(\tau) C_-^{(l,l)}(\tau)\nonumber \\
& \!\!+\! G^{(l,\bar{l})}\!(\tau) C_-^{(l,\bar{l})}\!(\tau)\Big]\!\!-\! i \!\! \int_{0}^{+\infty}\!\!\!\!\!\!\!\! \d \tau \,{\cal L}_{1,{\rm a}}(\tau) \!\Big[G^{(l,l)}\!(\tau) C_+^{(l,l)}\!(\tau)  \nonumber \\
& \!\!+\! G^{(l,\bar{l})}\!(\tau)C_+^{(l,\bar{l})}\!(\tau)\Big] \!\!+\! G^{(l,\bar{l})}\!(0)\Gamma_1(0) C^{(l,\bar{l})}\!(0). 
\end{align}

Recalling equations \eqref{eq:ls}--\eqref{eq:la}, we introduce the retarded Green function of the fluctuating force for the $\nu=1$ bath
\begin{equation}
\chi_1(t) \equiv i \theta(t) \langle \left[\xi_1(t),\xi_1(0) \right]\rangle = 2 i \theta(t) {\cal L}_{1,{\rm a}}(t)
\label{eq:chi_nu_xi_t}
\end{equation}
being ${\cal L}_{1,{\rm a}}(t)=\frac{1}{2}\left( \langle \xi_1(t)\xi_1(0)\rangle - \langle \xi_1(0)\xi_1(t)\rangle \right)$. 
The Fourier transform of the symmetric, ${\cal L}_{1,{\rm s}}(t)$, and anti-symmetric, ${\cal L}_{1,{\rm a}}(t)$, parts of ${\cal L}_{1}(t)$  read, respectively,
\begin{equation}
{\cal L}_{1,{\rm s}}(\omega) = {\cal J}_1(\omega) \coth\left(\frac{\omega}{2T_1}\right), \quad 
{\cal L}_{1,{\rm a}}(\omega) = {\cal J}_1(\omega).
\label{eq:calL_sa_omega}
\end{equation}
Using equation \eqref{eq:chi_nu_xi_t} and \eqref{eq:calL_sa_omega}, it follows 
\begin{equation}
\chi_1''(\omega) = {\cal L}_{1,{\rm a}}(\omega) = {\cal J}_1(\omega).
\end{equation}
We introduce now a shifted response function
\begin{equation}
\zeta_1 (\omega) = \chi_1(\omega) - \chi_1(\omega =0).
\end{equation}
Using the relations introduced so far, it is possible to prove that real and imaginary part of the Fourier transform of the memory damping kernel $\gamma_1(t)$ in equation~\eqref{eq:gamma} are, respectively,
\begin{equation}
\gamma_1'(\omega) =  \frac{1}{m}\frac{\zeta_1''(\omega)}{\omega}, \quad
\gamma_1''(\omega)= -\frac{1}{m}\frac{\zeta_1' (\omega)}{\omega},
\end{equation}
which can be summarized into $\zeta_1(\omega) = i m \omega \gamma_1(\omega)$. For a Ohmic spectral density with a Drude-Lorentz cut-off, ${\cal J}_1(\omega)=m\omega\gamma_1/(1+\omega^2/\omega_c^2)$, as in the present setup, we have $\gamma_1'(\omega) = {\cal J}_1(\omega) / (m \omega)$ and $\gamma_1''(\omega) = {\cal J}_1(\omega) / (m \omega_c)$, from which
\begin{equation}
\zeta_1'(\omega) =  -\frac{\omega}{\omega_c}{\cal J}_1(\omega), \quad
\zeta_1''(\omega) = {\cal J}_1(\omega).
\end{equation}

Now, considering the monochromatic drives of equation~\eqref{eq:drives} and following similar steps as done in Ref.~\cite{carrega_arxiv} we finally arrive at
$ P_l=P_l^{(0)}+\delta P_l$, 
where
\begin{align}
& P_l^{(0)} = -\Omega \int_{- \infty}^{+ \infty} \frac{\d \omega}{4 \pi m}\mathcal{J}_1(\omega+\Omega) N(\omega,\Omega) \nonumber\\
& \quad\quad \times \left[ f_l^2(\Omega){\chi_{2}^{(l,l)}}''(\omega) + \cos(\phi) f_l(\Omega)f_{{\bar l}}(\omega){\chi_{2}^{(l,\bar{l})}}''(\omega)\right], \label{eq:PA0}\\
& \delta P_{A/B} = \mp \Omega f_A(\Omega)f_B(\Omega)\sin(\phi) \int_{- \infty}^{+ \infty} \frac{\d \omega}{4 \pi m} \nonumber\\
&\quad\quad \times \left[  \zeta_1''(\omega+\Omega) {\chi_{2}^{(A,B)}}'(\omega) \coth\left(\frac{\omega + \Omega}{2T_1}\right) \right.\nonumber\\
&\quad\quad \left. +\zeta_1'(\omega+\Omega)  {\chi_{2}^{(A,B)}}''(\omega) \coth\left(\frac{\omega}{2T_2}\right)
\right].
\end{align}
Notice that the total power is eventually given by $P=P_A^{(0)}+P_B^{(0)}$ and it is reported in the main text in equation~\eqref{eq:PtotAB}. The expressions for the average heat currents can be derived analogously and are reported in equation~\eqref{eq:j1exp}.
As a final remark we quote the expressions of the fluctuations for the separate power contributions obtained at the lowest perturbative order by following similar steps as outlined above:
\begin{align}
D_{P_l} =&  \Omega^2 f_l^2(\Omega)\int_{-\infty}^{+\infty} \frac{\d\omega}{4\pi m}{\cal J}_1(\omega+\Omega) N(\omega,\Omega)\nonumber\\ 
&\times \coth(\frac{\omega}{2T_2}-\frac{\omega+\Omega}{2T_1}){\chi^{(l,l)}_2}''(\omega).
\end{align}
Finally, fluctuations associated to the total power at the lowest perturbative order are given by
\begin{align}\label{eq:app:defDP0} 
D_{P}  = & \Omega^2 \!\int_{-\infty}^{+\infty} \!\!\!\frac{\d\omega}{4\pi m}{\cal J}_1(\omega \!+\! \Omega) N(\omega,\Omega) \coth\!\left(\!\frac{\omega}{2T_2}\!-\!\frac{\omega\!+\!\Omega}{2T_1}\!\right)\nonumber\\ 
&\times \!\bigg[f_A^2(\Omega){\chi_2^{(A,A)}}''\!(\omega) \!+\! f_B^2(\Omega){\chi_2^{(B,B)}}''\!(\omega) \nonumber\\
&+\! 2\cos(\phi)f_A(\Omega)f_B(\omega){\chi_2^{(A,B)}}''\!(\omega)\bigg].
\end{align}
Notice that this quantity, unlike the average total power, is not just the sum of the two subsystem contributions, but it also includes mixed term proportional to $\cos(\phi)$.

\section{Proof of $Q_P \geq 2$}
\label{app:QP_geq_2}
Here we will prove, by contradiction, that
\begin{equation}\label{eq:app:QPdef}
Q_P=\dot{S}\frac{D_{P}}{P^2}\geq 2\,,
\end{equation}
see equation~(\ref{eq:QP_def}). We start observing that equation~(\ref{eq:PtotAB}) can be rewritten as
\begin{equation}\label{eq:app:defP} 
P=-\Omega\int_{-\infty}^{+\infty}\frac{\mathrm{d}\omega}{4\pi m}\mathcal{J}_1(\omega+\Omega)N(\omega,\Omega)\chi_{\mathrm{eff}}''(\omega)\,,
\end{equation}
where we have introduced for compactness
\begin{align}
\chi_{\mathrm{eff}}''(\omega)=& f_A^2(\Omega){\chi_2^{(A,A)}}''\!(\omega) \!+\!f_B^2(\Omega){\chi_2^{(B,B)}}''\!(\omega)\nonumber \\
&+2\cos(\phi)f_A(\Omega)f_B(\Omega){\chi_2^{(A,B)}}''\,,
\end{align}
and we can similarly rewrite the expression of the total power fluctuations of equation~\eqref{eq:app:defDP0}.
Finally, the entropy production rate $\dot{S}= - J_1/T_1 + (P+J_2)/T_2$ using equations~\eqref{eq:PtotAB}-\eqref{eq:j1exp} reads
\begin{equation}\label{eq:app:defS} 
\dot{S}=2\!\int_{-\infty}^{+\infty}\!\!\!\frac{\mathrm{d}\omega}{4\pi m}\left(\!\frac{\omega}{2T_2} \!-\! \frac{\omega\!+\!\Omega}{2T_1}\!\right)\!\mathcal{J}_1(\omega \!+\! \Omega) N(\omega,\Omega)\chi_{\mathrm{eff}}''(\omega).
\end{equation}
By plugging equations~(\ref{eq:app:defDP0}),~(\ref{eq:app:defP}) and (\ref{eq:app:defS})
into equation~(\ref{eq:app:QPdef}) and performing straightforward algebra, the following condition is found:
\begin{equation}\label{eq:app:Ilessthan0}
Q_P<2 \implies I=\int_{-\infty}^{+\infty}\!\!\int_{-\infty}^{+\infty}\!\!\mathrm{d}\omega\mathrm{d}\omega'\Phi_{\Omega}(\omega,\omega')<0\,,
\end{equation}
where
\begin{align}\label{eq:app:defPhi}
&\Phi_{\Omega}(\omega,\omega')=\chi_{\mathrm{eff}}''(\omega)\chi_{\mathrm{eff}}''(\omega')\left[\lambda_{\omega}\coth(\lambda_{\omega'})-1\right]\nonumber\\
&\times\mathcal{J}_1(\omega+\Omega)\mathcal{J}_1(\omega'+\Omega)N(\omega,\Omega)N(\omega',\Omega)\,,
\end{align}
and where $\lambda_{\omega}=\frac{\omega}{2T_2}-\frac{\omega+\Omega}{2T_1}$. We will now show that the condition of equation~(\ref{eq:app:Ilessthan0}) can never be satisfied.

To begin with it is convenient to rewrite $I$ in a more symmetric form by noting that one can also write 
$$I=\frac{1}{2}\int_{-\infty}^{+\infty}\int_{-\infty}^{+\infty}\mathrm{d}\omega\mathrm{d}\omega'\left[\Phi_{\Omega}(\omega,\omega')+\Phi_{\Omega}(\omega',\omega)\right]\,.
$$
Hence
\begin{align}\label{eq:app:Irewritten}
I&=\frac{1}{2}\int_{-\infty}^{+\infty}\int_{-\infty}^{+\infty}\mathrm{d}\omega\mathrm{d}\omega'G(\omega,\omega')\frac{N(\omega,\Omega)}{\lambda_{\omega}}\frac{N(\omega',\Omega)}{\lambda_{\omega'}}\nonumber\\
&\times\mathcal{J}_1(\omega+\Omega)\mathcal{J}_1(\omega'+\Omega)\chi_{\mathrm{eff}}''(\omega)\chi_{\mathrm{eff}}''(\omega')\,,
\end{align}
with
\begin{align}\label{eq:app:Iintermsofg} 
G(\omega,\omega')\!=\!\lambda_{\omega}^2\!\left[\lambda_{\omega'}\coth(\lambda_{\omega'})\right] \!+\! \lambda_{\omega'}^2\!\left[\lambda_{\omega}\coth(\lambda_{\omega})\right] \!-\! 2\lambda_{\omega}\lambda_{\omega'}.
\end{align}
Clearly, since $\lambda_{\omega}\coth(\lambda_{\omega})\geq1$ one has $G(\omega,\omega')\geq\left(\lambda_{\omega}-\lambda_{\omega'}\right)^2\geq 0$ and therefore the sign of equation~(\ref{eq:app:Irewritten}) is determined by the other factors in its integrand.

Now note that we can write
\begin{equation}\label{eq:app:Noverlambda}
\frac{N(\omega,\Omega)}{\lambda_{\omega}}=\frac{\coth(\frac{\omega+\Omega}{2T_1})\coth(\frac{\omega}{2T_2})-1}{\left(\frac{\omega+\Omega}{2T_1}-\frac{\omega}{2T_2}\right)\coth(\frac{\omega+\Omega}{2T_1}-\frac{\omega}{2T_2})}\,,
\end{equation}
where we have exploited the identity $$\frac{\coth(x)-\coth(y)}{y-x}=\frac{\coth(x)\coth(y)-1}{(x-y)\coth(x-y)}\,.$$
The denominator of equation~(\ref{eq:app:Noverlambda}) is strictly positive, while the identity 
\begin{equation}
\coth(x)\coth(y)-1=\frac{\cosh(x-y)}{\sinh(x)\sinh(y)}
\end{equation}
implies $\mathrm{sgn}\{\coth(x)\coth(y)-1\}=\mathrm{sgn}\{xy\}$, 
the hyperbolic cosine (sine) being an even (odd) function, 
where $\mathrm{sgn}\{x\}$ is the sign of $x$. Therefore
\begin{equation}
\mathrm{sgn}\left\{\frac{N(\omega,\Omega)}{\lambda_{\omega}}\right\}=\mathrm{sgn}\{\omega(\omega+\Omega)\}\,.
\end{equation}
One then concludes that the sign of the integrand in equation~(\ref{eq:app:Irewritten}) is given by
\begin{align}
&\mathrm{sgn}\left\{(\omega+\Omega)\mathcal{J}_1(\omega+\Omega)\right\}\mathrm{sgn}\left\{(\omega'+\Omega)\mathcal{J}_1(\omega'+\Omega)\right\}\nonumber\\
&\times\mathrm{sgn}\left\{\omega\chi_{\mathrm{eff}}''(\omega)\right\}\mathrm{sgn}\left\{\omega'\chi_{\mathrm{eff}}''(\omega')\right\}\,.
\end{align}
Since $\mathcal{J}_1(\omega)$ is an odd function with $\mathcal{J}_1(\omega)\geq0$ for $\omega\geq0$, one immediately sees that $\mathrm{sgn}\left\{(\omega+\Omega)\mathcal{J}_1(\omega+\Omega)\right\}\geq 0$.

Finally, we have 
\begin{equation}
\omega\chi_{\mathrm{eff}}''(\omega)=\gamma_2\omega^2\frac{\mathcal{N}(\omega)}{\left|\mathcal{D}(\omega)\right|^2}\,,
\end{equation} 
where 
\begin{align}
\mathcal{N}(\omega)=&f_B^2(\Omega)\left(\omega^2-\omega_A^2\right)+f_A^2(\Omega)\left(\omega^2-\omega_B^2\right)+\nonumber\\
+&2\cos(\phi)\prod_{l=A,B}f_l(\Omega)\left(\omega^2-\omega_l^2\right)
\end{align}
and where $\mathcal{D}(\omega)$ is given in equation~(\ref{eq:denominator2}). Here, $\mathcal{N}(\omega)$ is a quadratic form in $\omega^2$ with discriminant $-4f_A^2(\Omega)f_B^2(\Omega)(\omega_A^2-\omega_B^2)^2\sin^2(\phi)\leq0$, which proves that $\mathcal{N}(\omega)\geq0$, whence we conclude that $\omega\chi_{\mathrm{eff}}''(\omega)\geq0$. This finally shows that the integrand of equation~(\ref{eq:app:Irewritten}) is non-negative and thus $I\geq0$, contradicting the condition stated in equation~(\ref{eq:app:Ilessthan0}).
 Thus, it is proven that $Q_{P}\geq 2$.

\section{Asymptotic expressions}
\label{app_asym}
Here we report the expressions of the various power contributions in the two opposite regimes of small and large external frequency $\Omega$. These expressions are evaluated at phase $\phi=\pi/2$.

In the adiabatic regime $\Omega \ll \omega_l$ the total power expansion reads
\be
P \to \Omega^2 \sum_{l=A,B}f_l^2(\Omega)\alpha_l^{({\rm ad})},
\ee
with
\begin{equation} 
\alpha^{({\rm ad})}_l \!\! = \!-\!\!\! \int_{-\infty}^{+\infty}\!\!\!\!\frac{\d\omega}{4\pi m} \!\bigg[\! \dot{{\cal J}_1}(\omega) N(\omega,0) -\frac{ {\cal J}_1(\omega)}{2 T_1 \sinh^2 \! \big(\!\frac{\omega}{2T_1}\!\big)} \!\bigg] \!{\chi_{2}^{(l,l)}}''\!\!(\omega)
\end{equation}
where we used the notation $\dot{F}(\omega) = \d F(\omega)/\d\omega$. 
The subsystem power contributions instead become
\be
P_l \sim \delta P_l \to \Omega f_A(\Omega)f_B(\Omega){ \delta \alpha}^{({\rm ad})}_l,
\ee
where
\begin{align} 
\delta \alpha_{A/B}^{({\rm ad})} = &   \mp \int_{- \infty}^{+ \infty}\!\!\! \frac{\d \omega}{4 \pi m}
\bigg[ {\cal J}_1(\omega) {\chi_{2}^{(A,B)}}'\!(\omega) \coth\!\left(\! \frac{\omega\! }{2T_1} \!\right)  \nonumber\\
& - m\omega\gamma_1''(\omega) {\chi_{2}^{(A,B)}}''\!\!(\omega) \coth\!\left(\! \frac{\omega}{2T_2}\!\right)\!\!\bigg]. 
\end{align}
The associated fluctuations start as 
\begin{align}
D^{({\rm ad})}_{P_l} &= \Omega^2 f_l^2(\Omega)\int_{-\infty}^{+\infty}\frac{\d\omega}{4\pi m}{\cal J}_1(\omega)N(\omega,0)\nonumber \\
&\times \coth\left(\frac{\omega (T_1-T_2)}{2 T_1T_2}\right){\chi_2^{(l,l)}}''(\omega)~.
\end{align}

In the opposite diabatic regime $\Omega\gg \omega_l,\omega_c$, one gets
\begin{align}
P\to \Omega {\cal J}_1(\Omega)\sum_{l=A,B} f_l^2(\Omega)\alpha_l^{({\rm dia})},
\end{align}
where
\be 
\alpha_l^{({\rm dia})}=\int_{-\infty}^{+\infty}\frac{ \d \omega}{4\pi m}\coth\left(\frac{\omega}{2T_2}\right){\chi_2^{(l,l)}}''(\omega)~.
\ee
We also have 
\be 
\delta P_l \to \Omega f_A(\Omega)f_B(\Omega) {\cal J}_1 (\Omega) \left[\delta \alpha_l^{({\rm dia})}\frac{\Omega}{\omega_{c}} + \delta \beta^{({\rm dia})}_l\right]~,
\ee
with
\begin{align} 
\delta \alpha^{({\rm dia})}_{A/B} &=\pm\int_{-\infty}^{+\infty}\frac{\d \omega}{4\pi m}\coth\left(\frac{\omega}{2T_2}\right){\chi_2^{(A,B)}}''(\omega)~\nonumber \\
\delta \beta^{({\rm dia})}_{A/B} &=\mp \coth\left(\frac{\Omega}{2T_1}\right)\int_{-\infty}^{+\infty}\frac{\d \omega}{4\pi m}{\chi_2^{(A,B)}}'(\omega)~.
\end{align}
Notice that the last term $\delta \beta_l^{({\rm dia})}$ can be discarded when $\Omega \gg \omega_c$.
Finally, the associated fluctuations in the diabatic regime read 
\begin{align} 
D^{({\rm dia})}_{P_l} = & \Omega^2 f_l^2(\Omega){\cal J}_1(\Omega) \coth \left( \frac{\Omega}{2T_1} \right) \nonumber\\
&\times \int_{-\infty}^{+\infty} \frac{\d \omega}{4\pi m}\coth  \left( \frac{\omega}{2T_2} \right){\chi_2^{(l,l)}}''(\omega).
\end{align}

\section{TUR for the $l=B$ subsystem}
\label{app:pb}
\begin{figure}[!ht]
	\centering
	\includegraphics[width=\linewidth]{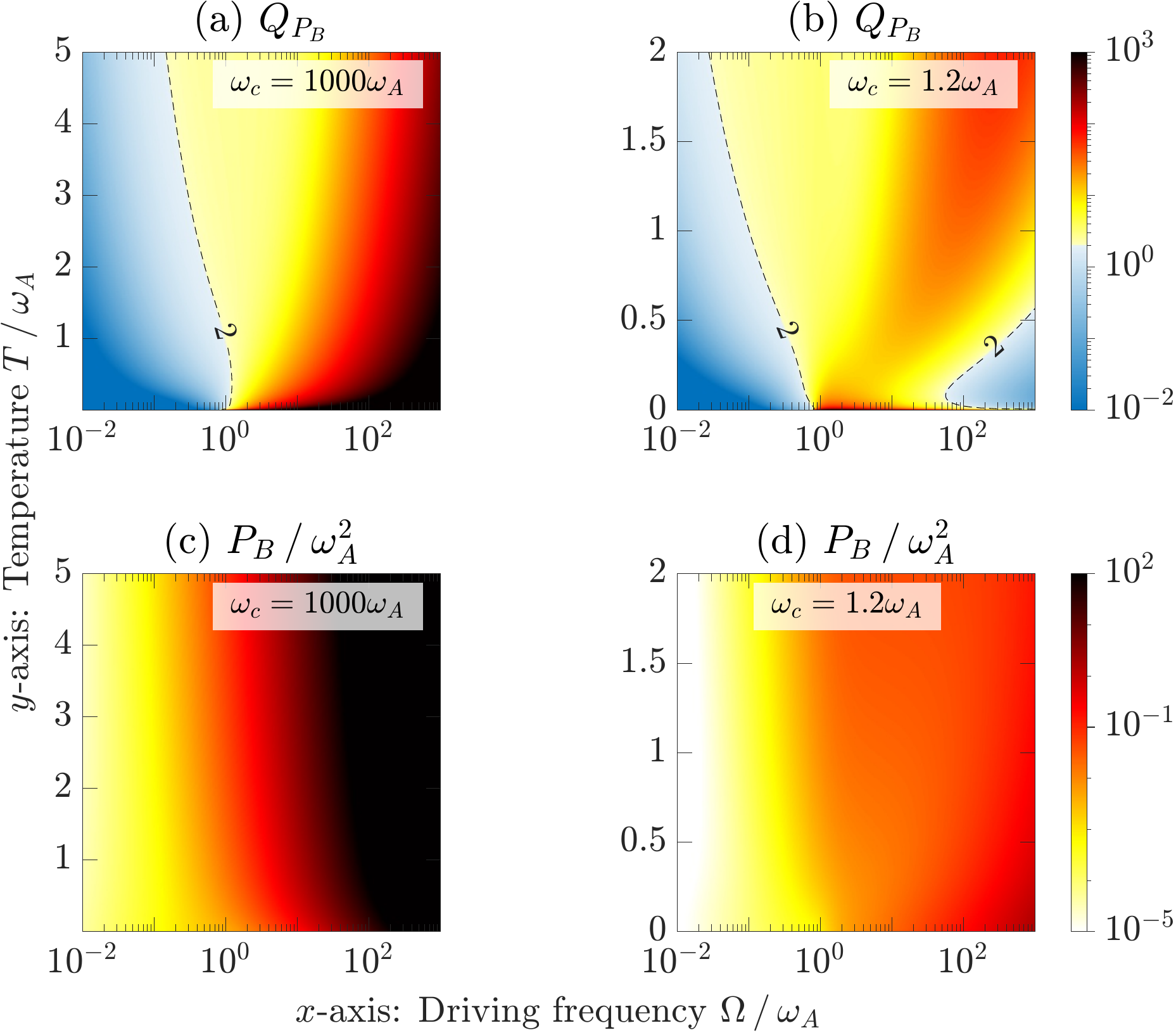}
	\caption{\textbf{Quantifying fluctuations of $P_B$ via the trade-off parameter $Q_{P_B}$  in the isothermal regime.} Density plots of $Q_{P_B}$ \textbf{(a)} and associated power $P_B$ \textbf{(c)} as a function of $\Omega/\omega_A$ and $T/\omega_A$ in the large cut-off regime $\omega_{c}=1000\omega_A$ and for $r=0$. Panels \textbf{(b,d)} report the same quantities as in panels (a,c) respectively but for a small cut-off $\omega_{c}=1.2\omega_A$. All results have been obtained in the strong damping regime $\gamma_2 =100\omega_A$ and other parameters as in figure~\ref{fig:2}.
	\label{fig:qpb}}
\end{figure}

Here we report results concerning the thermodynamic uncertainty relation and the associated power for the $l=B$ subsystem. These results have been obtained in the same parameter regions as in figure~\ref{fig:2} of the main text. Also in this case regions of TUR violation are present both in the adiabatic and diabatic regime as for $l=A$ discussed in the main text. However,
to these regions always correspond a positive $P_B>0$ 
and hence no useful resource can  be obtained from $l=B$ in this case (see figure~\ref{fig:qpb}).

\section{Work-to-work conversion efficiency} 
\label{app:eta}
In this appendix we consider the work-to-work conversion efficiency in the isothermal regime when $P_A <0$. We remind its definition:
\be
\eta=\frac{|P_A|}{P_B}=\frac{|P_A^{(0)} + f_B(\Omega) \delta P_A|}{f_B^2(\Omega) P_B^{(0)} -f_B(\Omega)\delta P_A}~,
  \ee
where, we recall, $f_B(\Omega)=(\Omega/\omega_c)^r$.

\begin{figure}[!ht]
	\centering
	\includegraphics[width=\linewidth]{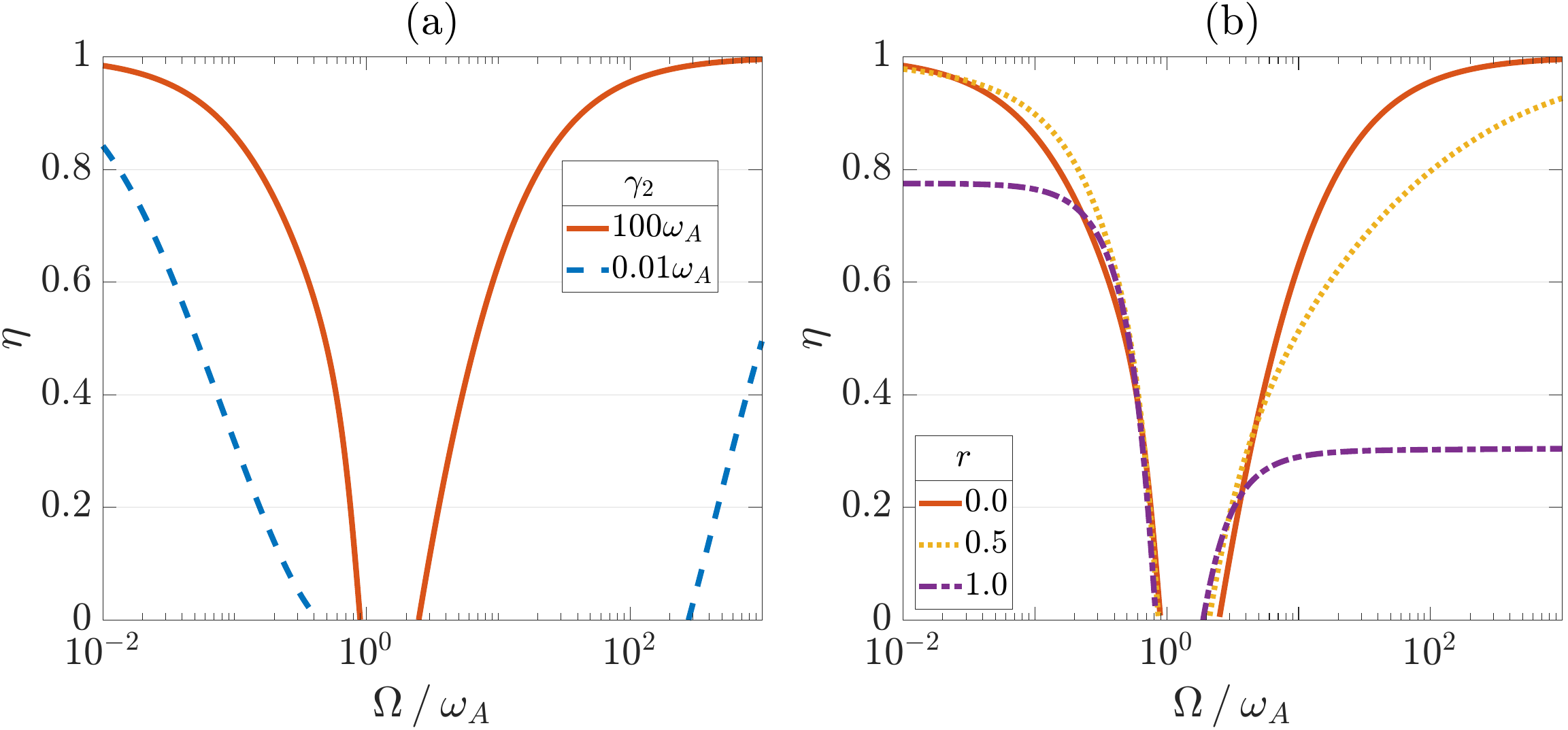}
	\caption{\textbf{Work-to-work conversion efficiency as a function of external frequency.}  
	\textbf{(a)} Work-to-work conversion efficiency in the weak ($\gamma_2 = 0.01\omega_A$) and strong ($\gamma_2 = 100\omega_A$) damping regime in the case $r=0$. Other parameters are $\gamma_1=0.001\omega_A$, $\omega_c=1.2\omega_A$, $T=0.1\omega_A$, and $\phi=\pi/2$. \textbf{(b)} Same quantity as in panel (a) but for different values of $r$ at strong damping ($\gamma_2=100\omega_A$).\label{fig:eta}}
\end{figure}

In figure~\ref{fig:eta}(a) we show the work-to-work conversion efficiency 
in the case of $r=0$ and in the two opposite case of weak and strong damping $\gamma_2$. This quantity is plotted as a function of the external frequency $\Omega$, showing that in the strong damping regime (when synchronization is established) optimal efficiency close to unity is achieved in the diabatic regime with sizeable output power.
In figure~\ref{fig:eta}(b), instead, we consider the work-to-work conversion efficiency at strong damping $\gamma_2=100\omega_A$ but for different values of the parameter $r$. 
Indeed we recall that both subsystem power in general will depend on the shape of $f_B(\Omega)=(\Omega/\omega_c)^r$.
In particular, in the diabatic regime one gets the asymptotic behaviour
\be
\eta\to \bigg[1 + \bigg(\frac{\Omega}{\omega_c}\bigg)^{r-1}\frac{\alpha_B^{{\rm dia}}}{|\delta \alpha_A^{{\rm dia}}|}\bigg]^{-1}
\ee
It is clear that if $r>1$ the efficiency in the diabatic regime tends to vanish, therefore one should consider $0\leq r < 1$ to still obtain a high efficiency.

\begin{figure}[!ht]
    \centering
	\includegraphics[width=\linewidth]{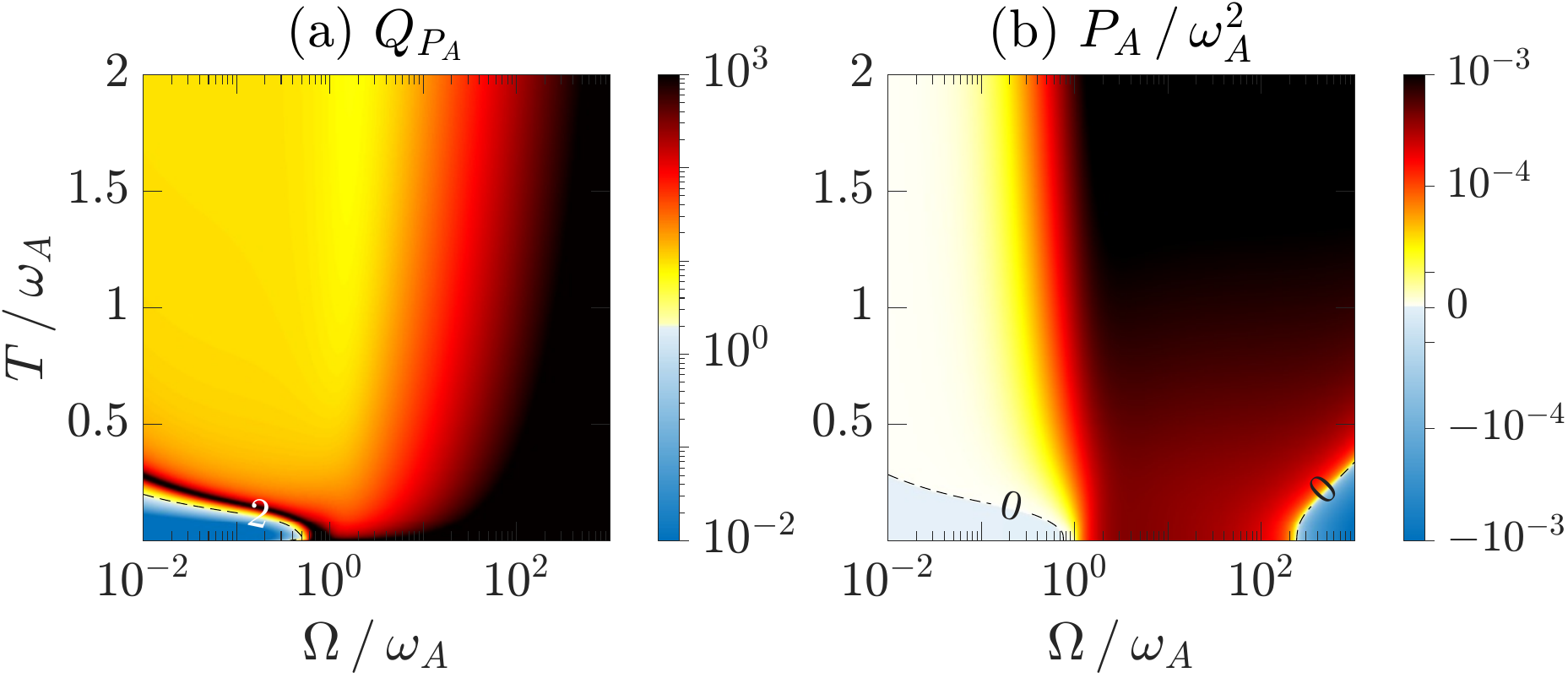}
	\caption{\textbf{Quantifying fluctuations of $P_A$ via the trade-off parameter $Q_{P_A}$ at weak damping.}
Density plots of $Q_{P_A}$ \textbf{(a)} and associated power $P_A$ \textbf{(b)} as a function of $\Omega/\omega_A$ and $T/\omega_A$ in the weak damping regime $\gamma_2=0.01 \omega_A$ and for $r=0$. Here $\gamma_1=0.001\omega_A$, the cut-off is $\omega_{c}=1.2\omega_A$. Other parameters as in figure~\ref{fig:2}.
	\label{fig:5}}
\end{figure}

\section{TUR at weak damping $\gamma_2$}
\label{app_weakdamping}
Here we present analogous results regarding the TUR quantifier $Q_{P_A}$ 
and the associated average power $P_A$ in the case of weak damping $\gamma_2\ll \omega_l$.

In particular, in figure~\ref{fig:5} we set $\gamma_2=0.01\omega_A$, and we show density plots for both quantities in the $\Omega-T$ plane.
As one can see, TUR violation in this case is present only in the low frequency and low temperature regime (left bottom corner of the plot in panel (a)). 
In this weak damping regime synchronization 
is absent and no TUR violation with sizeable power (in the diabatic regime) can be achieved.

\section{Frequency threshold for TUR violation at finite temperature gradient} 
\label{app_wth}
\begin{figure}[!ht]
    	\centering
 \includegraphics[width=\linewidth]{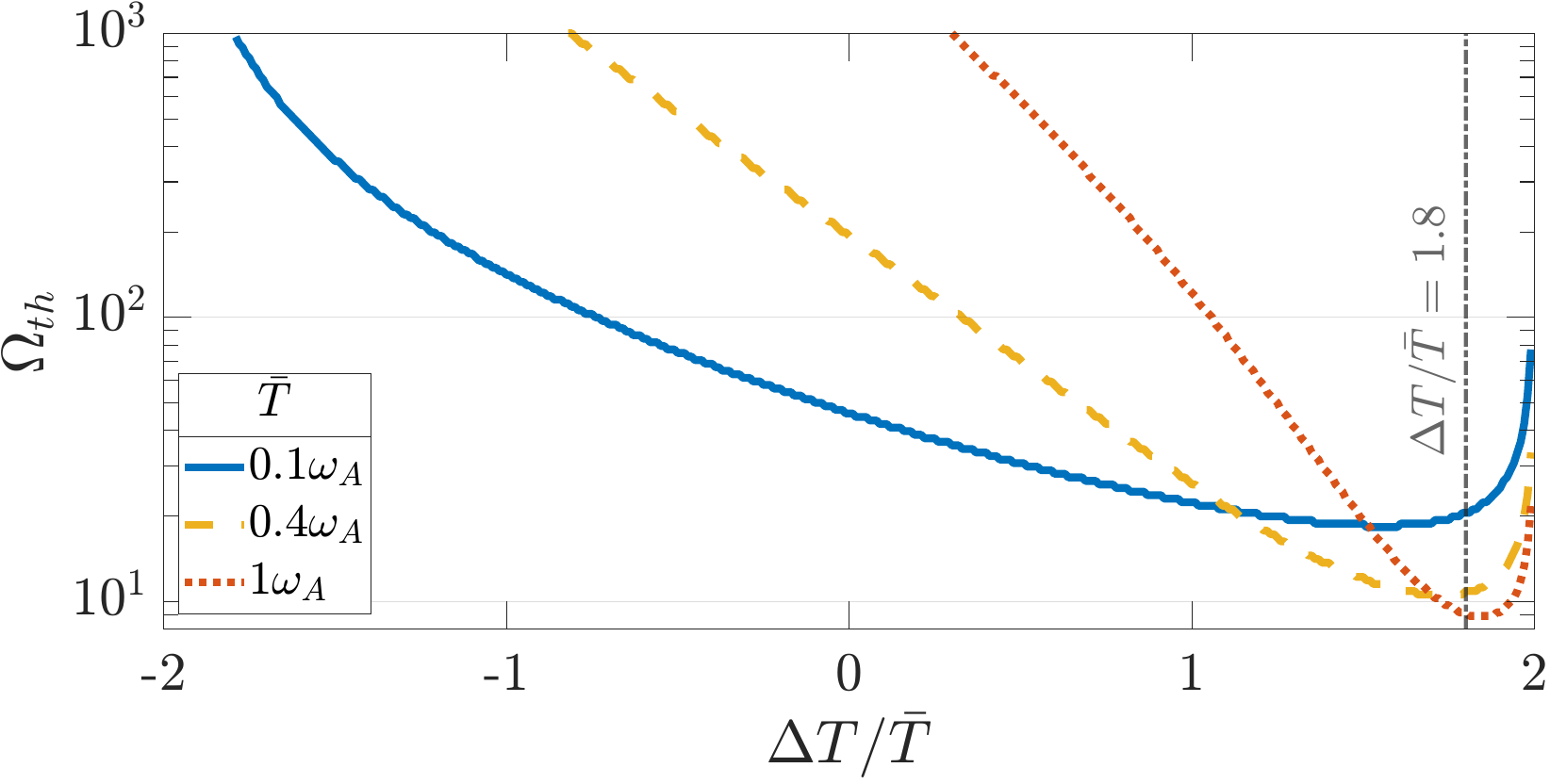}
\caption{\textbf{Frequency threshold for TUR violation at finite temperature gradient.}
Threshold driving frequency $\Omega_{\rm th}$ such that $Q_{P_A}<2$ for $\Omega > \Omega_{\rm th}$ as a function of $\Delta T/\bar{T}$ in the case $r=0$ and for different values of $\bar{T}$. 
Other parameters as in figure~\ref{fig:4}. 
\label{fig:wth}}
\end{figure}

In figure~\ref{fig:wth} we report the behaviour of the frequency threshold above which a TUR violation $Q_{P_A}<2$ is obtained in presence of a temperature gradient.
Here, one can see that a minimum of $\Omega_{\rm th}$ is attained at large positive values of $\Delta T / \bar{T}$ which lowers as $\bar{T}$ increases.

\end{document}